\documentclass[11pt, notitlepage]{article}

\usepackage{sectsty}
\allsectionsfont{\sffamily}
\usepackage{a4}
\usepackage{amssymb, graphicx,psfrag,epsf}
\usepackage{enumerate}
\usepackage{subfigure}
\usepackage{amsmath}
\usepackage{amsthm}
\usepackage{bbm}
\usepackage{undertilde}
\usepackage{verbatim}
\usepackage{bbm}
\usepackage{ dsfont }
\usepackage{geometry}
\usepackage{algorithm} 
\usepackage{algcompatible}
\usepackage{multirow}
\usepackage{pdflscape}
\usepackage{xcolor}
\usepackage{setspace}
\usepackage{hyperref}
\hypersetup{
	colorlinks=true,
	linkcolor=TUMblue,
	citecolor=TUMblue,
	filecolor=TUMblue,
	urlcolor=TUMblue
}
\usepackage{array}% http://ctan.org/pkg/array
\usepackage{xcolor}
\usepackage{bm}
\usepackage{cleveref}

\definecolor{TUMblue}{RGB}{0, 101, 189}
\definecolor{TUMlightblue}{RGB}{100,160,200}
\definecolor{TUMgreen}{RGB}{162,173,0}
\definecolor{TUMorange}{RGB}{227,114,034}
\definecolor{TUMivory}{RGB}{218,215,203}

\usepackage{aliascnt}

\newcommand{\mynewtheorem}[2]{
	\newaliascnt{#1}{dummy}
	\newtheorem{#1}[#1]{#2}
	\aliascntresetthe{#1}
	\expandafter\def\csname #1autorefname\endcsname{#2}
}

\theoremstyle{definition}
\mynewtheorem{thm}{Theorem}
\mynewtheorem{defi}{Definition}%[section]
\mynewtheorem{lem}{Lemma}%[section]
\mynewtheorem{cor}{Corollary}%[section]
\mynewtheorem{prop}{Proposition}%[section]
\mynewtheorem{exa}{Example}%[section]
\mynewtheorem{alg}{Algorithm}%[section]
\mynewtheorem{rem}{Remark}%[section]
\mynewtheorem{bsp}{Example}

\def\equationautorefname~#1\null{Equation~(#1)\null}

\newcolumntype{M}{>{\centering\arraybackslash}m{}}

\usepackage{url}
\usepackage{natbib}

\newcommand{\R}{\mathbb{R}}

\newcommand{\betab}{\boldsymbol{\beta}}

\newcommand{\lb}{\boldsymbol{l}}

\newcommand{\thetab}{\boldsymbol{\theta}}
\newcommand{\Ub}{\mathbf{U}}
\newcommand{\ub}{\mathbf{u}}

\newcommand{\vb}{\mathbf{v}}

\newcommand{\Xb}{\mathbf{X}}
\newcommand{\xb}{\mathbf{x}}

\newcommand{\yb}{\mathbf{y}}

\newcommand{\0}{\mathbf{0}}

\newcommand{\Fc}{\mathcal{F}}

\newcommand{\Nc}{\mathcal{N}}
\newcommand{\Mc}{\mathcal{M}}

\newcommand{\Uc}{\mathcal{U}}
\newcommand{\Xc}{\mathcal{X}}

\newcommand{\be}{\begin{equation}}
\newcommand{\ee}{\end{equation}}

\newcommand{\argmin}{\operatornamewithlimits{argmin}}
\newcommand{\argmax}{\operatornamewithlimits{argmax}}

\DeclareMathOperator{\AIC}{AIC}
\DeclareMathOperator{\BIC}{BIC}

\DeclareMathOperator{\cll}{cll}

\DeclareMathOperator{\MISE}{MISE}
\DeclareMathOperator{\RMISE}{RMISE}

\begin{document}

	\title{\bf D-vine copula based quantile regression}
	\date{\small \today}
	\author{Daniel Kraus\footnote{Corresponding author, Zentrum Mathematik, Technische Universit\"at M\"unchen, Boltzmanstra\ss e 3, 85748 Garching (email: \href{mailto:daniel.kraus@tum.de}{daniel.kraus@tum.de})}\hspace{.2cm} and Claudia Czado\footnote{Zentrum Mathematik, Technische Universit\"at M\"unchen, Boltzmanstra\ss e 3, 85748 Garching}}
	\maketitle

%\title{\textbf{D-vine copula based quantile regression
%}}
%\date{\small \today}
%\author{Daniel Kraus\\ \small{Technische Universit\"at M\"unchen}
%				\and Claudia Czado\\ \small{Technische Universit\"at M\"unchen}}
%
%\maketitle
\vspace{-3mm}
%\bigskip
  \begin{abstract}
%Quantile regression, that is the prediction of a random variable's quantiles conditioned on other random variables taking on certain values, has steadily gained importance in statistical modeling and financial applications.
Quantile regression, that is the prediction of conditional quantiles, has steadily gained importance in statistical modeling and financial applications. The authors introduce a new semiparametric quantile regression method based on sequentially fitting a likelihood optimal D-vine copula to given data resulting in highly flexible models with easily extractable conditional quantiles. As a subclass of regular vine copulas, D-vines enable the modeling of multivariate copulas in terms of bivariate building blocks, a so-called pair-copula construction (PCC). The proposed algorithm works fast and accurate even in high dimensions and incorporates an automatic variable selection by maximizing the conditional log-likelihood. Further, typical issues of quantile regression such as quantile crossing or transformations, interactions and collinearity of variables are automatically taken care of. In a simulation study the improved accuracy and saved computational time of the approach in comparison with established quantile regression methods is highlighted. An extensive financial application to international credit default swap (CDS) data including stress testing and Value-at-Risk (VaR) prediction demonstrates the usefulness of the proposed method.
  \end{abstract}

%\vfill
%\newpage
%---------------------------------------------------------------------------------------------------------------------------------------------------
%\spacingset{1.45}
\section{Introduction}\label{sec:introduction}

Predicting quantiles (e.g.\ median or quartiles) of a random variable conditioned on other variables taking on fixed values, has continually attracted interest and found applications in various fields, especially in finance. It has become a standard tool for risk managers working on portfolio optimization, asset pricing and the evaluation of systemic risk. For example, \cite{adrian2011covar} introduce the \emph{CoVaR}, a measure for systemic risk calculating conditional quantiles of a financial institution's loss distribution conditional on other institutions being in distress, and use it to evaluate the institution's contribution to systemic risk. A similar approach to measure systemic risk is found in \cite{brownlees2012volatility}. Further applications of quantile regression in the financial sector include measuring dependence in the FX markets \citep{bouye2009dynamic}, developing pricing models for real estates \citep{li2013optimal} and predicting volatilities in the stock market \citep{noh2014semiparametric}.\\
The literature is quite rich in methods to predict conditional quantiles. The most famous and therefore frequently used method is linear quantile regression \citep{koenker1978regression} which can be seen as the expansion of the well known ordinary least squares estimation used to predict conditional means. These simple linear models have been refined to account for nonparametric effects via additive models \citep{koenker2011additive,fenske2012identifying}. Further methods include local quantile regression \citep{spokoiny2013local}, single-index quantile regression \citep{wu2010single}, semiparametric quantile regression \citep{noh2014semiparametric}, nonparametric quantile regression \citep{li2013optimal} and quantile regression for time series \cite[e.g.][]{chen2009copula,xiao2012conditional}. In the machine learning context, \cite{hwang2005simple} use support vector machines for conditional quantile estimation while random forests are utilized in \cite{meinshausen2006quantile}. Moreover, \cite{bouye2009dynamic} propose a general approach to nonlinear quantile regression with one predictor based on a copula function.

The linear quantile regression method by \cite{koenker1978regression} has been criticized by \cite{bernard2015conditional} for imposing too restrictive assumptions on the shape of the regression quantiles. They show that for normally distributed marginals the model is misspecified as soon as the underlying dependence structure between response and covariates deviates from a Gaussian copula. Further, the method suffers from issues like quantile crossing and from the typical pitfalls of linear models such as multicollinearity, selection and significance of covariates and the inclusion of interactions or transformed variables. 

In contrast, the methodology proposed in this paper makes no precise assumptions about the shape of the conditional quantiles. The dependence relationship between response and covariates is modeled flexibly using a parametric D-vine copula, a subclass of regular vine copulas which have enhanced statistical modeling since the publication of the seminal paper of \cite{aas2009pair}. Then, as we will show, the model's conditional quantiles can be extracted analytically without approximations or excessive computational effort. As is usual when working with copulas, we further gain from the added flexibility of separating marginal and dependence modeling.

One of the main contributions of this work is a new algorithm that sequentially fits such a regression D-vine copula to given copula data, exhibiting many desirable features. On the one hand, step by step, the algorithm adds covariates to the regression model with the objective of maximizing a conditional likelihood, i.e.\ the likelihood of the predictive model of the response given the covariates. On the other hand, an automatic variable selection is incorporated, meaning that the algorithm will stop adding covariates to the model as soon as none of the remaining covariates is able to significantly increase the model's conditional likelihood. This results in parsimonious and at the same time flexible models whose conditional quantiles may strongly deviate from linearity. Due to the model construction, quantile crossings do not occur. Thus, the resulting D-vine quantile regression is able to overcome all the shortcomings of classical linear quantile regression mentioned above and therefore adds a new (and as we will see competitive) approach to the existing research on quantile regression.

%In this paper we go beyond the existing methodologies and use the flexible class of D-vine copula models for the prediction of conditional quantiles. D-vines are a subclass of regular vine copulas which have enhanced statistical modeling since the publication of the seminal paper of \cite{aas2009pair}. Vines enable the modeling of multivariate copula densities in terms of bivariate building blocks resulting in versatile parametric models. Given a response and predictor variables, we fit a D-vine to the data that maximizes the conditional likelihood of the response given the predictors by sequentially adding covariates to the model until the remaining covariates are deemed as non-influential. Thus, the question of variable selection is answered automatically. As we will show, the use of D-vine copulas allows us to extract the desired conditional quantiles precisely and without computational effort in a way that naturally avoids quantile crossing.\\
The remainder of the paper is organized as follows. \autoref{sec:DVines} introduces the concept of D-vines while \autoref{sec:QRModel} describes how they can be used for the prediction of conditional quantiles. Further, we demonstrate the usefulness of the approach in a simulation study. The prediction performance of D-vine quantile regression and its competitor methods, which are discussed in \autoref{sec:estqreg}, is presented in \autoref{sec:resultsofss}. For the demonstration of the usefulness of our method we include a real data example in \autoref{sec:cdsapp} that contains financial applications of the D-vine quantile regression featuring stress testing and Value-at-Risk prediction. \autoref{sec:conclusion} gives conclusions and areas of future research.

%---------------------------------------------------------------------------------------------------------------------------------------------------
\section{D-vine copulas}\label{sec:DVines}

%%%%%%%%%%%%%%%%%%%%%%%%%%%%%%%%%
%%%%%   shortened version:  %%%%%
%%%%%%%%%%%%%%%%%%%%%%%%%%%%%%%%%

A $d$-dimensional \textit{copula} $C$ is a $d$-variate distribution function on the unit hypercube $[0,1]^d$ with uniform marginal distribution functions. Sklar's Theorem \citep{sklar1959fonctions} provides a link between multivariate distributions and their associated copulas. It states that for every multivariate random vector $\Xb=(X_1,\ldots,X_d)'\sim F$ with marginal distribution functions $F_1,\ldots,F_d$, there exists a copula $C$ associated with $\Xb$, such that $F(x_1,\ldots,x_d)=C(F_1(x_1),\ldots,F_d(x_d))$. This decomposition of the multivariate distribution into its margins and its associated copula is unique when $\Xb$ is absolutely continuous (which we will assume for the remainder of this paper). In that case, the density of $\Xb$ can be decomposed similarly: $f(x_1,\ldots,x_d)=c(F_1(x_1),\ldots,F_d(x_d))f_1(x_1)\cdot\ldots\cdot f_d(x_d)$, where $c(u_1,\ldots,u_d):=\frac{\partial^d}{\partial_1\cdots\partial_d}C(u_1,\ldots,u_d)$ is the copula density and $f_1,\ldots,f_d$ are the marginal densities.\\
If we are interested solely in the dependence structure of $\Xb$, we considered it on the so-called \textit{u-scale} (or copula scale) by applying the \textit{probability integral transform} (PIT) to its marginals: $U_j:=F_j(X_j)$, $j=1,\ldots,d$.
The $U_j$ are then uniformly distributed and their joint distribution function is the copula $C$ associated with $\Xb$. Refer to \cite{joe1997multivariate} and \cite{nelsen2007introduction} for a detailed examination of copulas including many examples of parametric copulas, especially bivariate \textit{pair-copulas}. Those are of special interest to us since they are the building blocks used for the pair-copula construction of D-vines.

For a random vector $\Xb$, a set $D\subset\left\{1,\ldots,d\right\}$ and $i,j\in\left\{1,\ldots,d\right\}\backslash D$ we use the following notations:
%\begin{defi}[Notations]
%Let $(X_1,\ldots,X_d)'$ be a random vector, $D\subset\left\{1,\ldots,d\right\}$ and $i,j\in\left\{1,\ldots,d\right\}\backslash D$.% and $i_1,\ldots,i_k\in\left\{1,\ldots,d\right\}\backslash\left\{i,j\right\}$.
\begin{enumerate}
	\item[(a)] $C_{X_i,X_j;\Xb_D}(\cdot,\cdot;\xb_D)$ denotes the copula associated with the conditional distribution of $(X_i,X_j)'$ given $\Xb_D=\xb_D$. We abbreviate this by $C_{ij;D}(\cdot,\cdot;\xb_D)$. Further, $c_{ij;D}(\cdot,\cdot;\xb_D)$ is the copula density corresponding to $C_{ij;D}(\cdot,\cdot;\xb_D)$.
	\item[(b)] $F_{X_i|\Xb_D}(\cdot,\cdot|\xb_D)$ denotes the conditional distribution of the random variable $X_i$ given $\Xb_D=\xb_D$. We use $F_{i|D}(\cdot,\cdot|\xb_D)$ as an abbreviation.
	\item[(c)] $C_{U_i|\Ub_D}(\cdot,\cdot|\ub_D)$ denotes the conditional distribution of the PIT random variable $U_i$ given $\Ub_D=\ub_D$. We abbreviate this by $C_{i|D}(\cdot,\cdot|\ub_D)$.
	%\item[(b)] $c_{i,j;i_1,\ldots,i_k}:=c_{i,j;i_1,\ldots,i_k}\left(F_{i|i_1,\ldots,i_k}\left(x_i|x_{i_1},\ldots,x_{i_k}\right),F_{j|i_1,\ldots,i_k}\left(x_j|x_{i_1},\ldots,x_{i_k}\right);x_{i_1},\ldots,x_{i_k}\right)$.
	%	\item[(d)] Further, we define $c_{i,j;D}:=c_{X_i,X_j;\Xb_D}\left(F_{i|D}\left(x_i|\xb_D\right),F_{j|D}\left(x_j|\xb_D\right);\xb_D\right)$, where $c_{X_i,X_j;\Xb_D}(\cdot)$ is the copula density corresponding to $C_{X_i,X_j;\Xb_D}(\cdot)$.
\end{enumerate}

Following \cite{czado2010pair}, the joint density $f$ of the continuously distributed random vector $\Xb$ can be written in terms of (conditional) bivariate copula densities and its marginal densities as
\begin{multline}
	f(x_1,\ldots,x_d) = \prod_{k=1}^d{f_k(x_k)}\prod_{i=1}^{d-1}\prod_{j=i+1}^dc_{ij;i+1,\ldots,j-1}\big(F_{i|i+1,\ldots,j-1}\left(x_i|x_{i+1},\ldots,x_{j-1}\right),\\
	F_{j|i+1,\ldots,j-1}\left(x_j|x_{i+1},\ldots,x_{j-1}\right);x_{i+1},\ldots,x_{j-1}\big).
	\label{eq:D-vine_density}
\end{multline}
We call this pair-copula construction (PCC) a \emph{D-vine density} with order $X_1$--$X_2$--$\ldots$--$X_d$. % (see Figure \ref{fig:d_vine_numbers}). 
If all margins are uniform, we speak of a \emph{D-vine copula}. As introduced by \cite{bedford2002vines} we present a graph theoretic representation of the D-vine, where each edge of the graph corresponds to a pair-copula.
\begin{exa}\label{exa:5dim}
	\autoref{fig:d_vine_numbers} shows an exemplary 5-dimensional D-vine corresponding with
	\begin{align}
		f(x_1,x_2,x_3,x_4,x_5)=&f_1(x_1)f_2(x_2)f_3(x_3)f_4(x_4)f_5(x_5)\nonumber\\
		&\cdot c_{12}\cdot c_{23}\cdot c_{34}\cdot c_{45}\tag{$T_1$}\\
		&\cdot c_{13;2}\cdot c_{24;3}\cdot c_{35;4}\tag{$T_2$}\\
		&\cdot c_{14;23}\cdot c_{25;34}\tag{$T_3$}\\
		&\cdot c_{15;234},\tag{$T_4$}
	\end{align}
	where for brevity we omitted the arguments of the pair-copulas.
	\begin{figure}[htbp]
		\centering
		\includegraphics[width=.7\textwidth]{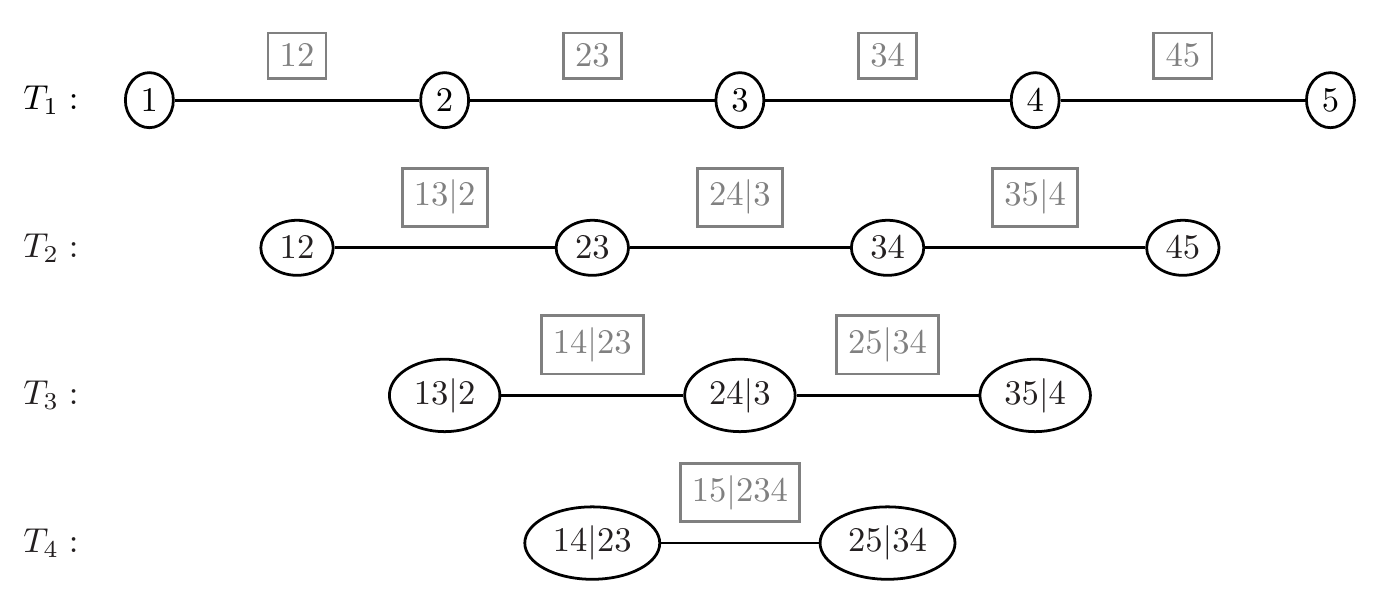}
		\caption{Graph theoretic representation of a D-vine with order $X_1$--$X_2$--$X_3$--$X_4$--$X_5$. The nodes of the trees are plotted in black circles and the corresponding pair-copulas in gray squares.}
		\label{fig:d_vine_numbers}
	\end{figure}
	
	%A common assumption when working with D-vine copulas is to assume that the copulas associated to conditional distributions $c_{i,j;i_1,\ldots,i_k}$ do not depend on the specific values of the conditioning vector $(x_{i_1},\ldots,x_{i_k})'$, i.e. $c_{i,j;i_1,\ldots,i_k}(\cdot,\cdot;x_{i_1},\ldots,x_{i_k})\equiv c_{i,j;i_1,\ldots,i_k}(\cdot,\cdot)$. Nevertheless, the arguments of the copulas, $F_{i|i_1,\ldots,i_k}\left(x_i|x_{i_1},\ldots,x_{i_k}\right)$ and $F_{j|i_1,\ldots,i_k}\left(x_j|x_{i_1},\ldots,x_{i_k}\right)$, obviously do depend on the conditioning values. A more detailed examination of this so-called simplifying assumption can be found in \cite{stober2013simplified}.\\
	We see that all pair-copulas used in the decomposition appear as edges in the corresponding nested set of trees displayed in \autoref{fig:d_vine_numbers}.
\end{exa}
In order to fit a D-vine copula with a fixed order to given data, all pair-copulas appearing in \autoref{eq:D-vine_density} are estimated as parametric bivariate copulas. A common assumption when working with D-vines is to assume that the copulas associated with conditional distributions $c_{i,j;D}$ do not depend on the specific values of the conditioning vector $\xb_D$, i.e.\ $c_{i,j;D}(\cdot,\cdot;\xb_D)\equiv c_{i,j;D}(\cdot,\cdot)$. A more detailed examination of this so-called simplifying assumption can for example be found in \cite{stober2013simplified}, \cite{haff2010simplified}, \cite{killiches2016examination} and \cite{killiches2016using}.\\
%The appearing conditional distributions $F_{i|i_1,\ldots,i_k}$ can be evaluated using only the pair-copulas specified for the D-vine by applying the following recursion, which was first stated by \cite{joe1997multivariate}: For $l\leq k$ it holds
%\begin{equation}
%F_{i|i_1,\ldots,i_k}\left(x_i|x_{i_1},\ldots,x_{i_k}\right) = \frac{\partial C_{i,i_l;i_1,\ldots,i_{l-1},i_{l+1},\ldots,i_k}\left(\mu,\nu\right)}{\partial \nu},
%\label{eq:Joe97}
%\end{equation}
%evaluated at 
%\[\mu = F_{i|i_1,\ldots,i_{l-1},i_{l+1},\ldots,i_k}\left(x_i|x_{i_1},\ldots,x_{i_{l-1}},x_{i_{l+1}},\ldots,x_{i_k}\right)\]
%and
%\[\nu=F_{i_l|i_1,\ldots,i_{l-1},i_{l+1},\ldots,i_k}\left(x_{i_l}|x_{i_1},\ldots,x_{i_{l-1}},x_{i_{l+1}},\ldots,x_{i_k}\right).\]
The conditional distributions $F_{i|D}\left(x_i|\xb_D\right)$ appearing in the PCC can be evaluated using only the pair-copulas specified for the D-vine from lower trees by applying the following recursion, which was first stated by \cite{joe1997multivariate}: Let $l\in D$ and $D_{-l}:=D\backslash \left\{l\right\}$. Then,
\begin{equation}
	F_{i|D}\left(x_i|\xb_D\right) = h_{i|l;D_{-l}}\left(F_{i|D_{-l}}\left(x_i|\xb_{D_{-l}}\right)|F_{l|D_{-l}}\left(x_{l}|\xb_{D_{-l}}\right)\right),
	\label{eq:Joe97}
\end{equation}
where for $i,j\notin D, i<j$, $h_{i|j;D}(u|v)={\partial C_{ij;D}(u,v)}/{\partial v}=C_{i|j;D}(u|v)$ and analogously $h_{j|i;D}(u|v)={\partial C_{ij;D}(u,v)}/{\partial u}=C_{j|i;D}(u|v)$ are the h-functions associated with the pair-copula $C_{ij;D}$.\\
%\begin{equation}
%F_{i|D}\left(x_i|\xb_D\right) = \frac{\partial C_{i,l;D_{-l}}\left(F_{i|D_{-l}}\left(x_i|\xb_{D_{-l}}\right),\nu\right)}{\partial \nu}\Big|_{\nu=F_{l|D_{-l}}\left(x_{l}|\xb_{D_{-l}}\right)}.
%\label{eq:Joe97}
%\end{equation}
In \autoref{exa:5dim} the first argument of $c_{14;23}$ from Tree 3, namely $F_{1|23}(x_1|x_2,x_3)$, can be evaluated using the h-functions associated with $C_{13;2}$, $C_{12}$ and $C_{23}$ from the first two trees:
%\begin{align*}
%	F_{1|23}(x_1|x_2,x_3) &= h_{1|3;2}(F_{1|2}(x_1|x_2)|F_{3|2}(x_3|x_2))\\
%	&=h_{1|3;2}(h_{1|2}(F_1(x_1)|F_2(x_2))|h_{3|2}(F_3(x_3)|F_2(x_2))).
%\end{align*}
\[
F_{1|23}(x_1|x_2,x_3)= h_{1|3;2}(F_{1|2}(x_1|x_2)|F_{3|2}(x_3|x_2))=h_{1|3;2}(h_{1|2}(F_1(x_1)|F_2(x_2))|h_{3|2}(F_3(x_3)|F_2(x_2))).
\]

\section{D-vine based quantile regression model}\label{sec:QRModel}

\subsection{Conditional quantile function}

The main purpose of D-vine copula based quantile regression is to predict the quantile of a response variable $Y$ given the outcome of some predictor variables $X_1,\ldots,X_d$, $d\geq 1$, where $Y\sim F_Y$ and $X_j\sim F_j$, $j=1,\ldots,d$. Hence, the focus of interest lies on the joint modeling of $Y$ and $\Xb$ and in particular on the conditional quantile function for $\alpha\in\left(0,1\right)$:
\begin{equation}
q_{\alpha}(x_1,\ldots,x_d):=F^{-1}_{Y|X_1,\ldots,X_d}(\alpha|x_1,\ldots,x_d).
\label{eq:defi_q_alpha}
\end{equation}
Using the probability integral transforms $V:=F_Y(Y)$ and $U_j:=F_j(X_j)$ with corresponding PIT values $v:=F_Y(y)$ and $u_j:=F_j(x_j)$, it follows that
\begin{align*}
F_{Y|X_{1},\ldots, X_{d}}(y|x_{1},\ldots,x_{d})&=P(Y\leq y|X_1=x_{1},\ldots,X_d=x_{d})\\
&=P(F_Y(Y)\leq v|F_1(X_1)=u_{1},\ldots,F_d(X_d)=u_{d})\\
&=C_{V|U_{1},\ldots, U_{d}}(v|u_{1},\ldots,u_{d}).
\end{align*}
Therefore, inversion yields
\begin{equation}
F^{-1}_{Y|X_1,\ldots,X_d}(\alpha|x_1,\ldots,x_d)=F^{-1}_Y\left(C^{-1}_{V|U_{1},\ldots, U_{d}}(\alpha|u_{1},\ldots,u_{d})\right).
\label{eq:cond_quantile}
\end{equation}
Hence, the conditional quantile function can be expressed in terms of the inverse marginal distribution function $F^{-1}_Y$ of the response $Y$ and the \textit{conditional copula quantile function} $C^{-1}_{V|U_{1},\ldots, U_{d}}$ conditioned on the PIT values of $\xb$. Note that this result was already stated for the one-dimensional predictor case in Equation (2) of \cite{bernard2015conditional}.\\ 
Now, we can obtain an estimate of the conditional quantile function by estimating the marginals $F_Y$ and $F_j$, $j=1,\ldots,d$, as well as the copula $C_{V,U_1,\ldots,U_d}$ and plugging them into \autoref{eq:cond_quantile}:
\begin{equation}
\hat q_{\alpha}(x_1,\ldots,x_d):=\hat F^{-1}_Y\left(\hat C^{-1}_{V|U_{1},\ldots, U_{d}}(\alpha|\hat u_{1},\ldots,\hat u_{d}%;\hat\thetab
)\right),
\label{eq:cond_quantile_est}
\end{equation}
where $\hat u_j:=\hat F_j(x_j)$ is the estimated PIT of $x_j$, $j=1,\ldots,d$%, and $\hat\thetab$ is the parameter vector of the estimated copula $\hat C_{V,U_{1},\ldots, U_{d}}$
.\\
While regarding the $\hat F_j$ there is a vast literature about the estimation of a univariate CDF, the question arises how to estimate the multivariate copula $C_{V,U_1,\ldots,U_d}$, such that on the one hand it facilitates a flexible model that is able to capture asymmetric dependencies, heavy tails and tail dependencies between the variables, and on the other hand the estimated conditional quantile function $\hat C^{-1}_{V|U_{1},\ldots, U_{d}}(\alpha|\hat u_{1},\ldots,\hat u_{d}%;\hat\thetab
)$ is easily calculable. The answer we suggest is to fit a D-vine copula to $(V,U_1,\ldots,U_d)'$, such that $V$ is the first node in the first tree (i.e.\ a D-vine with order $V$--$U_{l_1}$--$\ldots$--$U_{l_d}$, where $(l_1,\ldots,l_d)'$ is allowed to be an arbitrary permutation of $(1,\ldots,d)'$). This results in a flexible class of copulas since each bivariate copula of the pair-copula construction can be modeled separately and the order of the $U_j$ is a parameter that can be chosen such that the conditional likelihood is maximized as will be explained in detail in the next section. Finally, the recursion given in \autoref{eq:Joe97} allows us to express $C_{V|U_{1},\ldots, U_{d}}(v| u_{1},\ldots, u_{d})$ in terms of nested h-functions and consequently, $C^{-1}_{V|U_{1},\ldots, U_{d}}(\alpha| u_{1},\ldots, u_{d})$ in terms of inverse h-functions. We will now demonstrate this in a 4-dimensional example.

\begin{exa}
For a D-vine with order $V$--$U_{1}$--$U_{2}$--$U_{3}$, using \autoref{eq:Joe97} the conditional distribution of $V$ given $(U_1,U_2,U_3)'$ can recursively be expressed as 
\begin{align*}
C_{V|U_1,U_2,U_3}(v|u_1,u_2,u_3)&=h_{V|U_3;U_1,U_2}(C_{V|U_1,U_2}(v|u_1,u_2)|C_{U_3|U_1,U_2}(u_3|u_1,u_2))\\
&=h_{V|U_3;U_1,U_2}(h_{V|U_2;U_1}(C_{V|U_1}(v|u_1)|C_{U_2|U_1}(u_2|u_1))|\\
&\qquad\qquad\qquad h_{U_3|U_1;U_2}(C_{U_3|U_2}(u_3|u_2)|C_{U_1|U_2}(u_1|u_2)))\\
&=h_{V|U_3;U_1,U_2}(h_{V|U_2;U_1}(h_{V|U_1}(v|u_1)|h_{U_2|U_1}(u_2|u_1))|\\
&\qquad\qquad\qquad h_{U_3|U_1;U_2}(h_{U_3|U_2}(u_3|u_2)|h_{U_1|U_2}(u_1|u_2))).
\end{align*}
Inversion yields the conditional quantile function:
\begin{equation*}
\begin{split}
&C^{-1}_{V|U_1,U_2,U_3}(\alpha|u_1,u_2,u_3)=\\
&h^{-1}_{V|U_1}\left[h^{-1}_{V|U_2;U_1}\left\{h^{-1}_{V|U_3;U_1,U_2}\Big(\alpha\big|h_{U_3|U_1;U_2}\big(h_{U_3|U_2}(u_3|u_2)\big|h_{U_1|U_2}(u_1|u_2)\big)\Big)\Big|h_{U_2|U_1}(u_2|u_1)\right\}\Big|u_1\right].
\end{split}
\end{equation*}
\end{exa}
Note that $C^{-1}_{V|U_{1},\ldots, U_{d}}(\alpha| u_{1},\ldots, u_{d})$ is monotonically increasing in $\alpha$. Therefore, a crossing of quantile functions corresponding to different quantile levels is not possible. This issue of quantile crossing often arises in linear and non-linear quantile regression \cite[e.g.\ see the application section of][]{fenske2012identifying}. \cite{bernard2015conditional} show that in linear regression quantile functions may cross if non-Gaussian data is modeled. In addition, in (non-)linear quantile regression a substantial amount of effort has to be put into dealing with issues such as transforming response and covariates, including interactions among covariates and avoiding collinearity between covariates. Our approach solves these issues automatically since the distribution class given by the D-vines is much more flexible and makes less restrictive model assumptions how the covariates influence the response. This is also noted for regular vine regression by \cite{cooke2015vine}.
\subsection{Estimation process}

\begin{figure}[htbp]
	\centering
	\includegraphics[width=.9\textwidth]{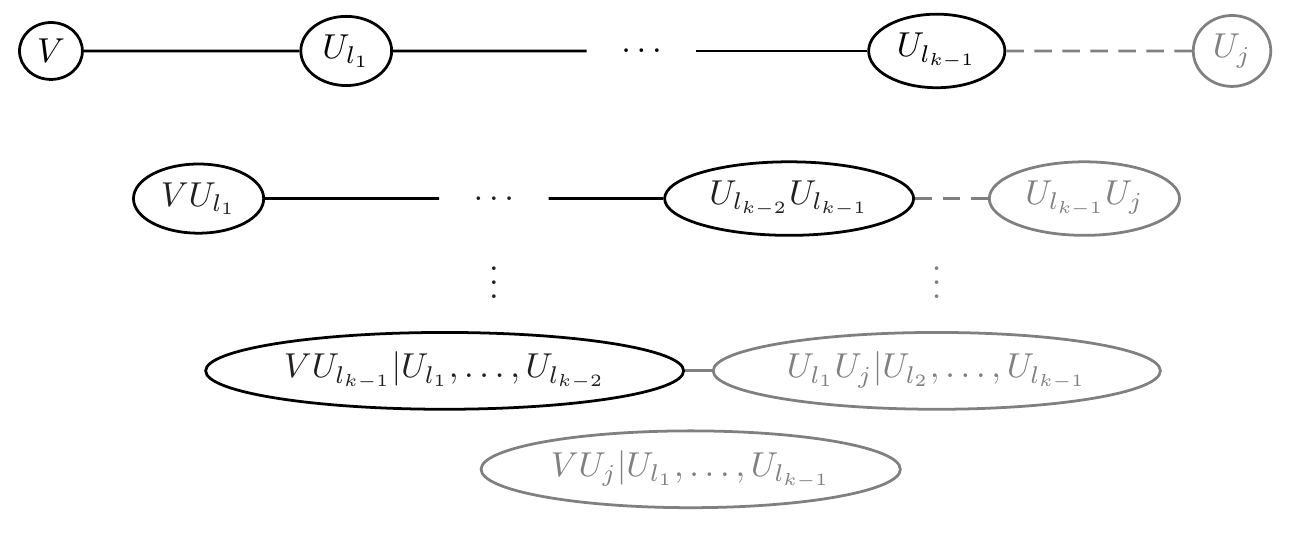}
	\caption{Extending the current D-vine (black) by adding $U_j$ in the $k$-th step of the algorithm. For this purpose, the gray pair-copulas have to be estimated.}
	\label{fig:Algo_Step_k}
\end{figure}

This section explains how the estimate of $q_{\alpha}(\xb)$ is obtained using a two step estimation procedure. All numerical calculations are done using the programming language R \citep{R} using the package \texttt{VineCopula} \citep{VineCopula}.\\
Let $\yb:=\left(y^{(i)}\right)_{i=1,\ldots,n}$, $\Xc:=\big(x^{(i)}_{j}\big)_{j=1,\ldots,d,\ i=1,\ldots,n}$ be $n$ independent and identically distributed observations of the random vector $\left(Y,X_1,X_2,\ldots,X_d\right)'$. The representation of $\hat q_{\alpha}(\xb)$ in \autoref{eq:cond_quantile_est} allows us to divide the estimation process into two steps. In the first step we estimate the marginal distribution functions $F_Y$ and $F_j$ of $Y$ and $X_j$, $j=1,\ldots,d$, respectively, and in the second step the D-vine that specifies the pair copulas needed to evaluate $\hat C^{-1}_{V|U_{1},\ldots, U_{d}}(\alpha|\hat u_{1},\ldots,\hat u_{d}%;\hat\thetab
)$ is estimated.
\paragraph{First step: estimation of the marginals}\mbox{}\\
In general, we have two choices of how to fit the marginal distributions, either parametrically or nonparametrically. Since we will fit the copula in the second step parametrically, this choice will either result in a fully parametric or semiparametric estimate of $q_{\alpha}(\xb)$. \cite{noh2013copula} point out that modeling the marginals as well as the copula parametrically might cause the resulting fully parametric estimator to be biased and inconsistent if one of the parametric models is misspecified. Therefore we prefer the semiparametric approach and estimate the marginals nonparametrically. Since we later need the inverse of the estimated marginals for the quantile prediction (c.f.\ \autoref{eq:cond_quantile_est}) we do not want to use the discrete valued ECDF for the estimation. Thus we choose the continuous kernel smoothing estimator \citep{parzen1962estimation}, which is, given a sample $\left(x^{(i)}\right)_{i=1,\ldots,n}$, defined as
\begin{equation}
\hat{F}(x)=\frac{1}{n}\sum_{i=1}^n{K\left(\frac{x-x^{(i)}}{h}\right)},\ x\in\R,
\label{eq:KSE}
\end{equation}
where $K(x):=\int_{-\infty}^{x}k(t)dt$ with $k(\cdot)$ being a symmetric probability density function and $h>0$ a bandwidth parameter. Usually, we choose $k=\varphi$, i.e.\ a Gaussian kernel, and the plug-in bandwidth developed in \cite{Duong2015}, Equation (4), which minimizes the asymptotic mean integrated squared error (implemented in the function \texttt{kcde} of the package \texttt{ks} \citep{Rks}).\\
%Silverman's ``rule of thumb'' (see \cite{silverman1986density}, page 48, equation (3.31)).
%\begin{defi}[Rescaled empirical cumulative distribution function]
%Let $\left(x^{(1)},\ldots,x^{(n)}\right)'$ be an i.i.d. sample of the random variable $X\sim F$. Then, the rescaled empirical cumulative distribution function of $X$ is defined as
%\[
%\hat{F}(x)=\frac{1}{n+1}\sum_{i=1}^n{I\left(x_i\leq x\right)},
%\]
%for all $x\in\R$, where $I(\cdot)$ denotes the indicator function.
%\end{defi}
Hence, we obtain $\hat{F}_Y$ and $\hat{F}_j$ as estimates for the marginal distribution functions. We use these to transform the observed data to pseudo copula data $\hat v^{(i)}:=\hat{F}_Y\left(y^{(i)}\right)$ and $\hat u^{(i)}_{j}:=\hat{F}_j\big(x^{(i)}_{j}\big)$, $j=1,\ldots,d$, $i=1,\ldots,n$. The pseudo copula data $\hat\vb=\left(\hat v^{(i)}\right)_{i=1,\ldots,n}$, $\hat\Uc=\big(\hat u^{(i)}_{j}\big)_{j=1,\ldots,d,\ i=1,\ldots,n}$ is then an approximately i.i.d.\ sample from the PIT random vector $(V,U_1,\ldots,U_d)'$ and will be used to estimate the D-vine copula in the second step.
%\subsubsection{Probability integral transform}\mbox{}\\
%In order to estimate the copula of a multivariate random vector, the marginals of the data have to be approximately uniformly distributed. We achieve this, by applying the probability integral transform to each of the marginals.
%\begin{defi}[Probability integral transform]
%Let $x_1,\ldots,x_n$ be an i.i.d. sample of the random variable $X\sim F$. Then, the probability integral transform (PIT) of $\xb$ is defined as
%\[
%\ub:=\left(F(x_1),\ldots,F(x_n)\right).
%\]
%\end{defi}
%\begin{rem}
%The probability integral transform of the random variable $X$, $U=F(X)$, is uniformly distributed. Hence, $\ub$ is a sample of a uniformly distributed random variable.
%\end{rem}
%Since the true distribution functions of $Y$ and the $X_j$ are unknown, we apply the PIT using the estimated distribution functions:\\
%\begin{align*}
%v^{(i)}&:=\hat{F}_Y(y^{(i)}),\\
%u^{(i)}_{j}&:=\hat{F}_i(x^{(i)}_{j}),\ j=1,\ldots,d,\ i=1,\ldots,n.
%\end{align*}
%The so called pseudo copula data $\vb=(v^{(i)})_{i=1,\ldots,n}$, $U=(u^{(i)}_{j})_{j=1,\ldots,d,\ i=1,\ldots,n}$ then has approximately uniformly distributed margins while still having the same dependence structure as the original sample $\left(\yb,X\right)$. This data will be used to fit the copula in the second step.
\paragraph{Second step: estimation of the D-vine}\mbox{}\\
%As motivated by Equation \eqref{eq:cond_quantile_est}, we fit a D-vine with order $V$--$U_{l_1}$--$\ldots$--$U_{l_d}$ to the pseudo copula data, because then all pair copulas needed to evaluate $\hat C^{-1}_{V|U_{1},\ldots, U_{d}}(\alpha|\hat u_{1},\ldots,\hat u_{d};\thetab)$ are estimated in the process. The ordering $l_1,\ldots,l_d$ is chosen in a likelihood optimal way according to the steps of Algorithm \ref{alg:D_Vine_seq} (see Appendix \ref{ch:Algo}).\\
%The main objective of the algorithm is to sequentially construct a D-vine while maximizing the model's conditional log-likelihood ($\cll$) in each step, where the $\cll$ of an estimated copula density $\hat c_{V,\Ub}(\cdot,\cdot;\thetab)$ given pseudo copula data is defined as 
%\[\cll\left(\hat c_{V,\Ub}(\cdot,\cdot;\thetab);\hat\vb,\hat\Uc\right) := \sum_{i=1}^n{\log\hat c_{V|\Ub}(\hat v^{(i)}|\hat u^{(i)}_{1},\ldots,u^{(i)}_{d};\thetab)}.
%\]
%We also consider two variants of the $\cll$ penalizing the number of parameters used
As motivated by \autoref{eq:cond_quantile_est}, we fit a D-vine with order $V$--$U_{l_1}$--$\ldots$--$U_{l_d}$ to the pseudo copula data, since then the evaluation of $\hat C^{-1}_{V|U_{1},\ldots, U_{d}}(\alpha|\hat u_{1},\ldots,\hat u_{d}%;\hat\thetab
)$ which is needed to calculate the conditional quantile is easily feasible. For this to work the ordering $\lb=(l_1,\ldots,l_d)'$ can generally be chosen arbitrarily. However, since the explanatory power of the resulting model does depend on the particular ordering, we want to choose it such that the resulting model for the prediction of the conditional quantile has the highest explanatory power. Since it would be infeasible to compare all $d!$ possible orderings, we propose a new algorithm that automatically constructs the D-vine sequentially choosing the most influential covariates. Similar to the \texttt{step} function for sequential estimation of linear models \cite[cf.][]{ripley2002modern}, starting with zero covariates, in each step we add the covariate to the model that improves the model's fit the most. As a measure for the model's fit, we define the conditional log-likelihood ($\cll$) of an estimated D-vine copula with ordering $\lb$, estimated parametric pair-copula families $\hat{\bm{\Fc}}$ and corresponding copula parameters $\hat\thetab$ %($=\frac{\partial^{d+1}}{\partial_1\cdots\partial_{d+1}}\hat C_{V,\Ub}(v,\ub;\hat\thetab)$) 
given pseudo copula data $(\hat\vb,\hat\Uc)$ as 
\begin{equation}
    \cll\left(\lb,\hat{\bm{\Fc}},\hat\thetab;\hat\vb,\hat\Uc\right) := \sum_{i=1}^n{\ln c_{V|\Ub}\left(\hat v^{(i)}|\hat \ub^{(i)};\lb,\hat{\bm{\Fc}},\hat\thetab\right)}.
\end{equation}
The conditional copula density $c_{V|\Ub}$ can be expressed as the product over all pair-copulas of the D-vine that contain $V$ \citep[see][]{killiches2015model}:  
\begin{multline*}
 c_{V|\Ub}\left(\hat v^{(i)}|\hat \ub^{(i)};\lb,\hat{\bm{\Fc}},\hat\thetab\right)=c_{VU_{l_1}}(\hat v^{(i)},\hat u_{l_1}^{(i)};\hat\Fc_{VU_{l_1}},\hat\theta_{VU_{l_1}})\times\\
\prod_{j=2}^{d} c_{VU_{l_j};U_{l_1},\ldots,U_{l_{j-1}}}\Big(\hat C_{V|U_{l_1},\ldots,U_{l_{j-1}}}\big(\hat v^{(i)}|\hat u_{l_1}^{(i)},\ldots,\hat u_{l_{j-1}}^{(i)}\big),\hat C_{U_{l_j}|U_{l_1},\ldots,U_{l_{j-1}}}\big(\hat u_{l_j}^{(i)}|\hat u_{l_1}^{(i)},\ldots,\hat u_{l_{j-1}}^{(i)}\big);\\
\hat\Fc_{VU_{l_j};U_{l_1},\ldots,U_{l_{j-1}}},\hat\theta_{VU_{l_j};U_{l_1},\ldots,U_{l_{j-1}}}\Big),
\end{multline*}
where $\hat\Fc_{I}$ and $\hat\theta_{I}$ denote the estimated family and parameter(s) of pair-copula $c_I$.\\
We now describe the D-vine regression algorithm which sequentially constructs a D-vine while maximizing the model's conditional log-likelihood in each step (for a detailed code see \autoref{ch:Algo}). Assume that at the beginning of the $k$th step of the algorithm the current optimal D-vine contains $k-1$ predictors (for illustration, see the black D-vine in \autoref{fig:Algo_Step_k}). For each of the remaining variables $U_j$ that have not been chosen yet, we fit the pair-copulas that are needed to extend the current model to a D-vine with order $V$--$U_{l_1}$--$\ldots$--$U_{l_{k-1}}$--$U_j$ (see the gray circles) and compute the resulting model's conditional log-likelihood. Then, the current model is updated by adding the variable corresponding to the highest $\cll$, concluding step $k$. That way, step by step the covariates are ordered regarding their power to predict the response.\\

%\begin{defi}[Conditional log-likelihood]
%Given pseudo copula data $\vb=(v^{(i)})_{i=1,\ldots,n}$, $U=(u^{(i)}_{j})_{j=1,\ldots,d,\ i=1,\ldots,n}$ and a conditional density $c_{V|\Ub}(\cdot,\cdot)$, the conditional log-likelihood is defined as
%\[
%\cll\left(c_{V|\Ub};\vb,U\right) = \sum_{i=1}^n{\log c_{V|\Ub}(v^{(i)}|u^{(i)}_{1},\ldots,u^{(i)}_{d})}.
%\]
%\label{defi:cll}
%\end{defi}
%Note that for a D-vine copula, $c_{V|\Ub}(v^{(i)}|u^{(i)}_{1},\ldots,u^{(i)}_{d})$ can be easily calculated using Lemma \ref{lem:cond_dens_mult_x_scale}.
In the case that in the $k$th step none of the remaining covariates is able to increase the model's $\cll$, the algorithm stops and returns the model only containing the $k-1$ chosen covariates so far. Therefore, an \textit{automatic forward covariate selection} is accomplished resulting in parsimonious models. In order to get even more parsimonious models, we also consider two variants of the $\cll$ penalizing the number of parameters $|\hat\thetab|$ used for the construction of the D-vine: the AIC-corrected conditional log-likelihood $\cll^{\AIC}$, defined as
%\[
%\cll^{\AIC}\left(\hat c_{V,\Ub}(\cdot,\cdot;\hat\thetab);\hat\vb,\hat\Uc\right):=\cll\left(\hat c_{V,\Ub}(\cdot,\cdot;\hat\thetab);\hat\vb,\hat\Uc\right)-|\hat\thetab|
%\]
\[
\cll^{\AIC}\left(\lb,\hat{\bm{\Fc}},\hat\thetab;\hat\vb,\hat\Uc\right):=-2\cll\left(\lb,\hat{\bm{\Fc}},\hat\thetab;\hat\vb,\hat\Uc\right)+2|\hat\thetab|
\]
and the BIC-corrected conditional log-likelihood $\cll^{\BIC}$, defined as
%\[
%\cll^{\BIC}\left(\hat c_{V,\Ub}(\cdot,\cdot;\hat\thetab);\hat\vb,\hat\Uc\right):=\cll\left(\hat c_{V,\Ub}(\cdot,\cdot;\hat\thetab);\hat\vb,\hat\Uc\right)-\frac{\log(n)}{2}|\hat\thetab|.
%\]
\[
\cll^{\BIC}\left(\lb,\hat{\bm{\Fc}},\hat\thetab;\hat\vb,\hat\Uc\right):=-2\cll\left(\lb,\hat{\bm{\Fc}},\hat\thetab;\hat\vb,\hat\Uc\right)+\log(n)|\hat\thetab|.
\]
Depending on how parsimonious the resulting model is desired to be, one can decide which version of the conditional log-likelihood to use. In our applications in the later sections we always use the AIC-corrected $\cll^{\AIC}$ since in the simulation study it has shown to select the most reasonable models in the sense that unimportant variables are disregarded and influential ones are kept in most of the instances.

\begin{exa}
We illustrate how the algorithm works for a four-dimensional data set $(y^{(i)},x^{(i)}_{1},x^{(i)}_{2},x^{(i)}_{3})'$, $i=1,\ldots,n=500$, sampled from $(Y,X_1,X_2,X_3)'\sim \Nc_4(\0,\Sigma)$ with 
\[\Sigma=\begin{pmatrix}
1 & 0.4 & 0.8 & 0 \\ 
0.4 & 1 & 0.32 & 0 \\ 
0.8 & 0.32 & 1 & 0 \\ 
0 & 0 & 0 & 1 \\
\end{pmatrix}.\]
First, the data is transformed to pseudo copula data $(\hat v^{(i)},\hat u^{(i)}_{1},\hat u^{(i)}_{2},\hat u^{(i)}_{3})'$, $i=1,\ldots,n$, using the kernel smoothing estimators introduced in \autoref{eq:KSE}.

In the \textbf{first step} of the algorithm, for each of the pairs $(V,U_j)'$, $j=1,2,3$, the AIC-optimal pair-copula is chosen using the function \texttt{BiCopSelect} of the R package \texttt{VineCopula} with an independence test at level 0.05 \citep[as described in][]{genest2007everything}. %(see \autoref{ch:Algo}, \autoref{alg:BiCopSelect})
Further, the conditional log-likelihood is calculated for each of the pairs (we omit the AIC- and BIC-corrected $\cll$-values here, since the fitted models have the same number of parameters and therefore these statistics would imply the same conclusions). The results are shown in the \autoref{tab:Algo_exa}. Implying the largest $\cll$, $U_2$ is chosen as the first variable to construct the D-vine.
\begin{table}[ht]
\centering
\begin{tabular}{|cccc|}
  \hline
Pair-copula & $\hat C_{V,U_1}$ & $\hat C_{V,U_2}$ & $\hat C_{V,U_3}$\\
  \hline
Family & Gauss & Gauss & Indep \\
Parameter & $0.34$ & $0.79$ & $0$\\
$\cll$ & $33.0$ & $249.4$ & $0$\\ 
%$\cll^{\AIC}$ & $32.0$ & $248.4$ & $0$\\ 
%$\cll^{\BIC}$ & $29.9$ & $246.3$ & $0$\\ 
   \hline
\end{tabular}
\caption{Candidate models with corresponding $\cll$ after the algorithm's first step.}
\label{tab:Algo_exa}
\end{table}

In the \textbf{second step}, we investigate whether the addition of either of the remaining variables $U_1$ or $U_3$ to the current D-vine can improve the conditional log-likelihood of the model. Adding $U_1$ would update the D-vine to order $V$--$U_2$--$U_1$ with newly estimated pair-copulas $\hat C_{V,U_1;U_2}$ (Gaussian with $\rho=0.27$) and $\hat C_{U_1,U_2}$ (Gaussian with $\rho=0.23$). The log-likelihood of the resulting conditional copula $\hat c_{V|U_2,U_1}$ is $269.8$. The addition of $U_3$ would result in both new pair-copulas to be estimated as independence copulas. Consequently, the conditional log-likelihood would not improve compared to the model without $U_3$. Since $269.8>249.4$, we update the vine to order $V$--$U_2$--$U_1$.

In the \textbf{third step}, we check whether the addition of the remaining variable $U_3$ to the D-vine improves the conditional log-likelihood of the model. Not surprisingly, as in the second step the new pair copulas $\hat C_{V,U_3;U_2,U_1}$, $\hat C_{U_2,U_3;U_1}$ and $\hat C_{U_1,U_3}$ are all estimated to be the independence copula. Hence, the conditional log-likelihood of the full model with order $V$--$U_2$--$U_1$--$U_3$ is equal to the one with order $V$--$U_2$--$U_1$. Consequently, the algorithm stops and returns the D-vine with order $V$--$U_2$--$U_1$.
%, which is displayed in Figure \ref{fig:Algo_Example3F}.
%\begin{figure}[htbp]
%	\centering
%	\includegraphics[width=0.41\textwidth]{Algo_Example3F.pdf}
%	\caption{D-vine returned by the algorithm. All pair-copulas were chosen to be Gaussian with $\rho$ equal to the numbers in brackets. $U_3$ was not chosen to be in the model.}
%	\label{fig:Algo_Example3F}
%\end{figure}
\end{exa}
This example demonstrates the main advantages of the proposed algorithm: It automatically selects the influential covariates, ranks them by their strength of predicting the response, disregards any superfluous variables and finally flexibly models the dependence between the response and the chosen covariates. Thus, the typical issues of regression such as collinearity, transformation and inclusion/exclusion of covariates are solved without any additional effort.

In the following simulation study, we compare the performance of D-vine based quantile regression to conventional and established quantile regression methods in settings where the true quantile is known.

\section{Established quantile regression methods}\label{sec:estqreg}
\paragraph{Linear quantile regression (LQR)}

\cite{koenker1978regression} were one of the first researchers in the field of quantile regression. For their linear quantile regression model, they assume the predicted conditional quantiles to be \textit{linear} in the predictors, i.e.\ 
\begin{equation}
	\hat q_{\alpha}\left(x_1^{(i)},\ldots,x_d^{(i)}\right):=\beta_0+\sum_{j=1}^d{\beta_j x_j^{(i)}},
\end{equation}
where the regression coefficients $\betab\in\R^{d+1}$ are chosen to solve the minimization problem
\begin{equation}\label{eq:LQR_min}
	\min_{\betab\in\R^{d+1}}\left\{ \alpha\sum_{i=1}^n{\left(y^{(i)}-\beta_0-\sum_{j=1}^d{\beta_j x_j^{(i)}}\right)^+}+(1-\alpha)\sum_{i=1}^n{\left(\beta_0+\sum_{j=1}^d{\beta_j x_j^{(i)}}-y^{(i)}\right)^+}\right\}.
\end{equation}
Linear quantile regression has drawn criticism for its strong assumption of linear conditional quantiles. \cite{bernard2015conditional} highlight its pitfalls showing that the linearity assumption is almost never fulfilled. For example, assuming normal margins the only copula resulting in linear conditional quantiles is the Gaussian copula, which is very restrictive. Another major flaw of linear quantile regression \cite{bernard2015conditional} point out is that regression lines of different quantile levels may cross, since they may have differing slopes. These drawbacks are also seen in the results of the simulation. Regarding implementation, we use the R-function \texttt{rq} of the package \texttt{quantreg} \citep{Rquantreg}.

\paragraph{Boosting additive quantile regression (BAQR)}

In order to relax the linearity assumption of the above method \cite{koenker2005quantile} proposes to use additive models for quantile regression, i.e.\
\begin{equation}
\hat q_{\alpha}\left(x_1^{(i)},\ldots,x_d^{(i)},z_1^{(i)},\ldots,z_J^{(i)}\right):=\beta_0+\sum_{j=1}^d{\beta_j x_j^{(i)}}+\sum_{j=1}^J{g_j\left(z_j^{(i)}\right)},
\end{equation}

where $g_j$ denotes a smooth function of the continuous variable $z_j^{(i)}$, $j=1,\ldots,J$. Commonly, these functions are estimated by means of B-spline basis functions with a rather small number of knots at fixed points, solving a minimization similar to \autoref{eq:LQR_min} with additional penalization of the total variation of the derivatives of the smooth functions \citep[see][]{koenker2011additive}. In contrast, \cite{fenske2012identifying} use boosting techniques for the estimation of additive quantile regression and show in simulation studies the superiority over the total variation regularization approach. In short, boosting minimizes a given loss
function (analogue to \autoref{eq:LQR_min}) by stepwise updating the estimator along the steepest gradient descent of the loss function. The optimal number of boosting iterations is determined as in \cite{fenske2012identifying} using a test data set. The algorithm is implemented in the function \texttt{gamboost} of the package \texttt{mboost} \citep{Rmboost}, for which we choose the parameter settings exactly as in the electronic supplement of \cite{fenske2012identifying}.

\paragraph{Nonparametric quantile regression (NPQR)}

A fully nonparametric estimation of a conditional quantile is proposed by \cite{li2013optimal}. After estimating the conditional distribution function $\hat F(y|x_1,\ldots,x_d)$ nonparametrically by applying a kernel estimator with an automatic data-driven bandwidth selector, the authors estimate the conditional quantile function $q_{\alpha}(x_1,\ldots,x_d)$ by numerically inverting the estimated conditional distribution function, solving
\begin{equation}
	\hat q_{\alpha}(x_1,\ldots,x_d):=\argmin_q\left|\alpha-\hat F(q|x_1,\ldots,x_d)\right|.
\end{equation}
In R, we use the function \texttt{npqreg} of package \texttt{np} \citep{Rnp}, where the bandwidths are selected automatically by the function \texttt{npcdistbw} (with default settings).

\paragraph{Semiparametric quantile regression (SPQR)}

The semiparametric method of \cite{noh2014semiparametric} uses a copula density weighted quantile regression approach. Similar to our procedure, \cite{noh2014semiparametric} propose a copula-based multivariate model for which they estimate the marginal distributions nonparametrically and the copula as a regular vine $\hat c$ \citep[see][]{aas2009pair}, which is a generalization of D-vines. Since in this case the conditional copula quantile function $C^{-1}_{V|U_{1},\ldots, U_{d}}$ in \autoref{eq:cond_quantile_est} can in general not be calculated as elegantly as in the case of a fitted D-vine, the authors suggest to estimate the conditional quantile via minimizing a weighted check function:
\begin{equation}\label{eq:SPQR_min}
	\hat q_{\alpha}(x_1,\ldots,x_d)=\argmin_a\sum_{i=1}^n{\rho_{\alpha}(y^{(i)}-a)\hat c\left(\hat{F}_Y(y^{(i)}),\hat F_1(x_1^{(i)}),\ldots,F_d(x_d^{(i)})\right)},
\end{equation}
where $\rho_{\alpha}(y)=y(\alpha-\mathbbm{1}(y<0))$ denotes the check function. Since this method is currently not implemented in R, we use the function \texttt{RVineStructureSelect} of the package \texttt{VineCopula} for the estimation of the regular vine and solve for the conditional quantile minimizing \autoref{eq:SPQR_min} using the \texttt{optimize} function.

While this approach might appear quite similar to our proposed one, the main difference between D-vine and semiparametric quantile regression is that our approach is able to analytically solve for the conditional quantiles while the semiparametric method computes the quantile by numerical minimization. This works especially badly when the sample size is small (and takes a very long time if it is large). Further, our approach estimates the model by selecting covariates with the objective of explaining the response (by maximizing the cll), while the method of \cite{noh2014semiparametric} estimates an R-vine that fits the overall distribution to all variables with no particular focus on the response. 

\section{Results of the simulation study}\label{sec:resultsofss}
\label{ch:ss}
We consider the following three scenarios for the random vector $(Y,X_1,\ldots,X_d)'$:

\begin{itemize}
	\item \textbf{C3}: $(Y,X_1,X_2)'$ follows a three-dimensional Clayton copula with parameter $\delta_1$ or $\delta_2$ and margin set $\mathcal{M}_1$ or $\mathcal{M}_2$ (see \autoref{tab:scenarios}, row 1).
	\item \textbf{t5}: $(Y,X_1,\ldots,X_4)'$ follows a five-dimensional t copula with 3 degrees of freedom, association matrix $R_1$ or $R_2$ and margin set $\mathcal{M}_1$ or $\mathcal{M}_2$ (see \autoref{tab:scenarios}, row 2).
	\item \textbf{M5}: $\Xb\sim N_4(\mathbf{0},\Sigma)$, with $\Sigma_{ij}=0.5^{|i-j|}$.\\
	$Y:=\sqrt{|2x_1-x_2+0.5|} + (-0.5x_3+1)(0.1x_4^3)+\sigma\epsilon$ with $\epsilon\sim N(0,1)$, $\sigma\in\left\{0.1,1\right\} $.
%	\item \textbf{N10}: $(Y,X_1,\ldots,X_9)'$ follows a ten-dimensional Gaussian copula with correlation matrix $R_1$ or $R_2$ and margin set $\mathcal{M}_1$ or $\mathcal{M}_2$ (see Table \ref{tab:scenarios}).
%	\item \textbf{M5}: $\Xb\sim N_6(\mathbf{0},\Sigma)$, with $\Sigma_{ij}=0.5^{|i-j|}$.\\
%	$Y:=((−2X_1 + X_2 - 4X_3 + 3X_4 + X_5 + 2X_6)/\sqrt{35})^3+\sigma\epsilon$ with $\epsilon\sim N(0,1)$, $\sigma>0$
\end{itemize}
%\begin{landscape}
\begin{table}[ht]
	{\footnotesize
	\begin{center}
		\tabcolsep=0.05cm
	\begin{tabular}{|>{\centering\arraybackslash}m{1.2cm}|>{\centering\arraybackslash}m{6.2cm}|>{\centering\arraybackslash}m{6.7cm}|}
		\hline
		Scen. & Copula parameter & Marginals\\
		\hline
\textbf{C3} &
\begin{tabular}[t]{l}  $\delta_1=0.86$ \\ 
	$\delta_2=4.67$ 
\end{tabular}
& \vspace{-1.9mm} \begin{tabular}[t]{|c|c|c|c|}
	\hline
	  & $Y$ & $X_1$ & $X_2$ \\ 
 	\hline 
 	$\Mc_1$ & $\Nc(0,1)$ & $t_4(0,1)$ & $\Nc(1,4)$ \\
 	\hline 
 	$\Mc_2$ & $st_4(0,1,2)$ & $s\Nc(-2,0.5,3)$ & $st_3(1,2,5)$ \\
 	\hline
 	    \end{tabular}
 	    \vspace{1mm} \\
		\hline
			\textbf{t5} & \vspace{1mm}
			\begin{tabular}[t]{l}  $R_1=\begin{pmatrix}
				1 & 0.6 & 0.5 & 0.5 & 0.4 \\ 
				0.6 & 1 & 0.5 & 0.5 & 0.5 \\ 
				0.5 & 0.5 & 1 & 0.5 & 0.5 \\ 
				0.5 & 0.5 & 0.5 & 1 & 0.5 \\ 
				0.4 & 0.5 & 0.5 & 0.5 & 1 \\ 
				\end{pmatrix}$ \vspace{1mm} \\ 
				$R_2=\begin{pmatrix}
				1 & 0.27 & 0.74 & 0.72 & 0.41 \\ 
				0.27 & 1 & 0.28 & 0.29 & 0.27 \\ 
				0.74 & 0.28 & 1 & 0.74 & 0.42 \\ 
				0.72 & 0.29 & 0.74 & 1 & 0.40 \\ 
				0.41 & 0.27 & 0.42 & 0.40 & 1 \\ 
				\end{pmatrix}$ 
			\end{tabular}\vspace{1mm}
			& \vspace{-3mm}
			\begin{tabular}[t]{l}
				\hspace{-3.4mm}
				\begin{tabular}[t]{|c|c|c|c|}\hline
					  & $Y$ & $X_1$ & $X_2$  \\ 
				\hline 
				$\Mc_1$ & $\Nc(0,1)$ & $t_4(0,1)$ & $\Nc(1,4)$  \\
				\hline 
 	$\Mc_2$ & $st_4(0,1,2)$ & $s\Nc(-2,0.5,3)$ & $st_3(1,2,5)$  \\
 	\hline
			\end{tabular}\\
			\hspace{-2.25mm}\begin{tabular}[t]{|c|c|c|}
				\hline
				  & $X_3$ & $X_4$ \\ 
				\hline 
				$\Mc_1$ & $t_4(0,1)$ & $\Nc(1,4)$ \\
				\hline 
				$\Mc_2$ & $s\Nc(-2,0.5,3)$ & $st_3(1,2,5)$\\
				\hline
				\end{tabular}
			\end{tabular}
			 \\
		\hline
	\end{tabular}
	\caption{Parameter and marginal settings for Scenarios \textbf{C3} and \textbf{t5} of the simulation study. $s\Nc(\mu,\sigma^2,\xi)$ and $st_{\nu}(\mu,\sigma^2,\xi)$ denote the skewed normal and skewed t distribution with $\nu$ degrees of freedom, respectively. $\xi$ is the skewness parameter as described in \cite{azzalinicollaboration}. Higher values of $\xi$ correspond to more skewed distributions.}
	\label{tab:scenarios}
\end{center}
}
\end{table}
In these scenarios the true quantile functions are known. See \autoref{sec:AnalyticalCondQuant} for an analytical derivation of the conditional quantiles of the multivariate t copula as well as the three-dimensional Clayton copula.\\
%\end{landscape}
To assess the performance of each of the estimation methods we consider the estimated out-of-sample mean integrated square error ($\hat{\MISE}_m$) of method $m$, defined as
%\[
%\hat{\MISE}:=\frac{1}{N}\sum_{l=1}^{N}\left[\frac{1}{I}\sum_{i=1}^{I}\left(\hat q^{(l)}_{\alpha}\left( \xb^{(i)}\right) -q_{\alpha}\left( \xb^{(i)}\right) \right)^2\right], 
%\]
\begin{equation}
\hat{\MISE}_m:=\frac{1}{R}\sum_{r=1}^{R}\left[\frac{1}{n_{eval}}\sum_{i=1}^{n_{eval}}\left\{\hat q^{(r)}_{m,\alpha}\left( \xb^{eval}_{r,i}\right) -q_{\alpha}\left( \xb^{eval}_{r,i}\right) \right\}^2\right], 
\end{equation}
%where, for each iteration $l=1,\ldots,N$ with $N=100$, $\hat q^{(l)}_{\alpha}(\cdot)$ is the estimator based on a training sample of size $n$, and $\xb^{(i)}$, $i=1,\ldots,I$ with $I=500$ is the evaluation sample simulated from the distribution of $\Xb$.\\
%For each of the four DGPs we present the performance of the estimation methods relative to the proposed D-vine based quantile regression (such that values greater than one imply a worse performance) for different training sample sizes $n$, quantile levels $\alpha$ and several copula and marginal parameter settings (see Appendix \ref{ch:res_sim_study}).\\
%Wee see that...
where, for each replication $r=1,\ldots,R=100$, we simulate a training data set $\left(y^{train}_{r,i},\xb^{train}_{r,i}\right)$, $i=1,\ldots,n_{train}$, from the distribution of $(Y,\Xb)'$, and further simulate an evaluation data set $\left(\xb^{eval}_{r,i}\right)$, $i=1,\ldots,n_{eval}$, of the distribution of $\Xb$. In this study we choose $n_{eval}=\frac12 n_{train}$. For each replication $r$ we estimate $\hat q^{(r)}_{m,\alpha}(\cdot)$, the quantile estimator of method $m$ based on the training data set. Then, we use the evaluation data set to estimate the integrated square error between the resulting estimates and the true quantiles. Finally, the mean over all replications is taken to yield the estimated mean integrated square error.\\
In order to better compare the performance of the competitor methods relative to D-vine regression we also consider the relative estimated mean integrated square error of method $m$, defined as
\begin{equation}
\hat{\RMISE}_m:=\frac{\hat{\MISE}_m}{\hat{\MISE}_{DVQR}},
\end{equation}
where $\hat{\MISE}_{DVQR}$ is the estimated MISE of D-vine quantile regression. Values greater than one imply a worse relative performance compared to D-vine quantile regression.\\
For each of the scenarios we present a table containing its results. The first two columns of each table specify the parameter settings for the margins and dependence structure. The third column gives information about the sample size of the training data set used to fit the models, while the fourth column determines which quantile is estimated. The fifth column contains the estimated MISE values obtained using D-vine quantile regression. The relative performances of the competitor methods can be found in columns six to nine in terms of the estimated RMISE as defined above. 
\paragraph{Results for Scenario C3} At first we will consider the results of Scenario \textbf{C3} in \autoref{tab:C3Comp}. A plot of the true quantile functions for this scenario is displayed in \autoref{fig:Clayton_quantiles} of \autoref{sec:AnalyticalCondQuant}.
\begin{table}[ht]
	\centering
	\small
	\begin{tabular}{|c|c|c|c||r|r|r|r|r|}
		\hline
		\multirow{2}{*}{margins} & \multirow{2}{*}{$\delta$} & \multirow{2}{*}{$n_{train}$} & \multirow{2}{*}{$\alpha$} & \multirow{2}{*}{$\hat{\MISE}_{DVQR}$}&\multicolumn{4}{c|}{$\hat{\RMISE}_m$}\\
		\cline{6-9}
		& & & & & LQR & BAQR & NPQR & SPQR\\
		\hline
		\multirow{8}{*}{$\Mc_1$} & \multirow{4}{*}{$\delta_1$} & \multirow{2}{*}{300} & 0.5 & 0.0118 & 4.08 & 2.87 & 4.39 & 0.97\\
		\cline{4-9}
		& & & 0.95&0.0252 & 3.94 & 3.13 & 4.23 & 1.09\\
		\cline{3-9}
		& & \multirow{2}{*}{1000} & 0.5&0.0036 & 12.91 & 5.26 & 7.56 & 1.07\\
		\cline{4-9}
		& & & 0.95&0.0083 & 7.02 & 4.20 & 7.28 & 1.09\\
		\cline{2-9}
		& \multirow{4}{*}{$\delta_2$}  & \multirow{2}{*}{300} & 0.5&0.0029 & 10.60 & 5.72 & 9.24 & 1.63\\
		\cline{4-9}
		& & & 0.95&0.0171 & 7.21 & 4.21 & 6.80 & 1.22\\
		\cline{3-9}
		& & \multirow{2}{*}{1000} & 0.5& 0.0011 & 30.90 & 11.86 & 16.80 & 1.46\\
		\cline{4-9} 
		& & & 0.95& 0.0054 & 20.06 & 7.09 & 12.90 & 1.19\\
		\cline{1-9}
		\multirow{8}{*}{$\Mc_2$} & \multirow{4}{*}{$\delta_1$} & \multirow{2}{*}{300} & 0.5& 0.0078 & 6.13 & 3.71 & 6.74 & 1.12\\
		\cline{4-9}
		& & & 0.95&0.0776 & 3.40 & 2.43 & 6.74 & 1.06\\
		\cline{3-9}
		& & \multirow{2}{*}{1000} & 0.5& 0.0022 & 20.53 & 6.02 & 14.55 & 1.19\\
		\cline{4-9}
		& & & 0.95& 0.0209 & 7.54 & 4.05 & 13.45 & 1.09\\
		\cline{2-9}
		& \multirow{4}{*}{$\delta_2$}  & \multirow{2}{*}{300} & 0.5& 0.0036 & 10.55 & 6.59 & 19.24 & 1.98\\
		\cline{4-9}
		& & & 0.95& 0.0664 & 3.29 & 2.43 & 8.76 & 1.17\\
		\cline{3-9}
		& & \multirow{2}{*}{1000} & 0.5& 0.0012 & 31.28 & 11.90 & 35.62 & 2.37\\
		\cline{4-9} 
		& & & 0.95& 0.0183 & 8.28 & 4.73 & 18.72 & 1.22\\
		\hline
	\end{tabular}
	\caption{$\hat{\MISE}$ of D-vine quantile regression and the relative performances of the other estimation methods based on data generated by Scenario \textbf{C3} with different parameter settings as specified in \autoref{tab:scenarios}.}
	\label{tab:C3Comp}
\end{table}

With regard to the estimated mean integrated square error of D-vine quantile regression (column 5) we see that the model fits very well with errors ranging in the order of magnitude $10^{-3}$ to $10^{-2}$ (we do not show plots of the fitted quantiles since they are almost identical to those displayed in \autoref{fig:Clayton_quantiles}). We generally observe that the 50\%-quantile has a better fit than the 95\%-quantile which is not surprising since the univariate estimator of the 50\%-quantile is more robust. Further, it is clear that the prediction errors are reduced when the training sample size increases. It is also plausible that the size of errors drops when changing from the setting of low dependence ($\delta_1$) to high dependence ($\delta_2$). Concerning the marginal distributions it is reasonable that the performance of the 95\%-quantile prediction worsens when the response distribution is skewed and heavy-tailed, since its marginal estimation is more imprecise, especially in the tails.\\
Comparing the estimated MISE values of D-vine quantile regression with the other methods described in \autoref{sec:estqreg} using the estimated relative mean integrated square error, we see that it outperforms its competitors by a great margin. The plots of the true conditional quantiles in \autoref{fig:Clayton_quantiles} imply that the linearity assumption of linear quantile regression is clearly violated in this scenario which results in severe prediction errors up to 30 times higher than our method. Using boosting additive models that allow for a nonlinear relationship manage to improve these results, however still lagging behind D-Vine quantile regression by a factor of 2.4 to 11.9. Further, in this example nonparametric quantile regression also seems too imprecise to be a serious competitor to D-vine quantile regression. Being conceptually closest to our method, it is not surprising that the semiparametric approach performs similarly well. However, except for one parameter setting it is still outperformed by the D-vine quantile regression considerably. Especially in the case of high dependence ($\delta_2$) and skewed margins ($\Mc_2$) its predictions generate notably higher errors. Possible explanations for this are given in the last paragraph about SPQR in \autoref{sec:estqreg}.

At this point we shortly focus on the computational times of the different quantile regression methods. Since they do not depend on the margins or copula parameter, we present in \autoref{tab:runtimes} the computational times for the exemplary setting $\delta_1$ and $\Mc_1$ for different sizes of the training data set. The times are given in seconds needed for the R=100 repetitions of estimating the 0.5- and 0.95-quantiles.
\begin{table}[ht]
	\centering
	\begin{tabular}{|r|r|r|r|r|r|}
		\hline
		$n_{train}$ & DVQR & LQR & BAQR & NPQR & SPQR \\ 
		\hline
		300 & 446 & 0.79 & 44 & 3065 & 4125 \\ 
		1000 & 1364 & 1.23 & 84 & 22768 & 24144 \\  
		\hline
	\end{tabular}
	\caption{Computational times in seconds for the different quantile regression methods in the exemplary setting $\delta_1$ and $\Mc_1$.}
	\label{tab:runtimes}
\end{table}

Due to the simplicity of linear quantile regression its computation times are unbeatable. Also, boosting additive models are considerably faster than the remaining methods. However, we observe that D-vine quantile regression is much faster than its nonparametric and semiparametric competitors. Further, the computational time of D-vine quantile regression grows linearly in $n_{train}$ while for the other two methods it seems to increase at a much higher rate. Moreover, we note that for the prediction of quantiles at different $\alpha$-levels D-vine quantile regression has the advantage of having to fit the model only once and then being able to easily extract different conditional quantiles by evaluating the inverse h-functions. For the other methods a separate optimization has to be performed for each quantile level.
\paragraph{Results for Scenario t5} The competitive advantage of D-vine quantile regression over the competitor methods is also supported by the results of Scenario \textbf{t5} displayed in \autoref{tab:t5Comp}.

\begin{table}[htbp]
	\centering
	\small
	\begin{tabular}{|c|c|c|c||r|r|r|r|r|}
		\hline
		\multirow{2}{*}{margins} & \multirow{2}{*}{$\delta$} & \multirow{2}{*}{$n_{train}$} & \multirow{2}{*}{$\alpha$} & \multirow{2}{*}{$\hat{\MISE}_{DVQR}$}&\multicolumn{4}{c|}{$\hat{\RMISE}_m$}\\
		\cline{6-9}
		& & & & & LQR & BAQR & NPQR & SPQR\\
		\hline
		\multirow{8}{*}{$\Mc_1$} & \multirow{4}{*}{$R_1$} & \multirow{2}{*}{300} & 0.5 & 0.0357 & 0.62 & 1.28 & 2.68 & 1.46\\
		\cline{4-9}
		& & & 0.95&0.0774 & 1.98 & 1.65 & 2.26 & 1.17\\
		\cline{3-9}
		& & \multirow{2}{*}{1000} & 0.5&0.0063 & 1.90 & 2.81 & 9.05 & 1.63\\
		\cline{4-9}
		& & & 0.95&0.0303 & 4.18 & 2.73 & 5.03 & 0.75\\
		\cline{2-9}
		& \multirow{4}{*}{$R_2$}  & \multirow{2}{*}{300} & 0.5&0.0214 & 0.59 & 1.27 & 2.45 & 1.23\\
		\cline{4-9}
		& & & 0.95&0.0738 & 1.46 & 1.36 & 1.87 & 0.97\\
		\cline{3-9}
		& & \multirow{2}{*}{1000} & 0.5&0.0052 & 1.34 & 2.33 & 6.56 & 2.18\\
		\cline{4-9} 
		& & & 0.95&0.0278 & 3.44 & 2.58 & 3.80 & 0.98\\
		\cline{1-9}
		\multirow{8}{*}{$\Mc_2$} & \multirow{4}{*}{$R_1$} & \multirow{2}{*}{300} & 0.5&0.0319 & 0.91 & 1.92 & 3.81 & 1.60\\
		\cline{4-9}
		& & & 0.95&0.2912 & 1.37 & 1.08 & 1.64 & 1.04\\
		\cline{3-9}
		& & \multirow{2}{*}{1000} & 0.5&0.0103 & 2.61 & 3.54 & 10.92 & 1.44\\
		\cline{4-9}
		& & & 0.95&0.1702 & 2.10 &  1.47 & 2.63 & 0.97\\
		\cline{2-9}
		& \multirow{4}{*}{$R_2$}  & \multirow{2}{*}{300} & 0.5&0.0250 & 0.81 & 1.61 & 4.08 & 1.53\\ 
		\cline{4-9}
		& & & 0.95&0.1842 & 1.17 & 1.02 & 1.71 & 1.06\\
		\cline{3-9}
		& & \multirow{2}{*}{1000} & 0.5&0.0078 & 1.90 & 2.64 & 12.74 & 2.13\\
		\cline{4-9} 
		& & & 0.95&0.0948 & 2.05 & 1.45 & 3.27 & 1.17\\
		\hline
	\end{tabular}
		\caption{$\hat{\MISE}$ of D-vine quantile regression and the relative performances of the other estimation methods based on data generated by Scenario \textbf{t5} with different parameter settings.}
	\label{tab:t5Comp}
\end{table}
 However, with the t-distribution being a little bit closer to the Gaussian model, the linearity assumption of the linear quantile regression method is not violated as severely resulting in smaller relative errors. In fact, in the setting of a small training sample size $n_{train}=300$, the prediction of the median is a bit more accurate than the prediction using any of the other methods. Nevertheless, in the tails of the distribution ($\alpha=0.95$), where the t- and the Gaussian distribution differ the most, it looses this property. Again, the results of the boosting additive and nonparametric method are well behind the others with especially bad relative performance in predicting the median when the sample size is large. Concerning the prediction of median values D-vine quantile regression also outperforms the semiparametric approach whose relative errors lie between 1.23 and 2.18. In the tail they perform quite similar with relative errors ranging between 0.75 and 1.17. All in all, D-vine quantile regression is still the preferred method, especially for larger sample sizes.
\paragraph{Results for Scenario M5} Finally, \autoref{tab:M5Comp} displays the results corresponding to Scenario \textbf{M5}.
\begin{table}[ht]
	\centering
	\small
	\begin{tabular}{|c|c|c||r|r|r|r|r|}
		\hline
		\multirow{2}{*}{$\sigma$} & \multirow{2}{*}{$n_{train}$} & \multirow{2}{*}{$\alpha$} & \multirow{2}{*}{$\hat{\MISE}_{DVQR}$}&\multicolumn{4}{c|}{$\hat{\RMISE}_m$}\\
		\cline{5-8}
		& & & & LQR & BAQR & NPQR & SPQR\\
		\hline
		\multirow{4}{*}{0.1} & \multirow{2}{*}{300} & 0.5&0.209 & 1.39 & 0.67 & 0.41 & 0.95\\
		\cline{3-8}
		& & 0.95&0.514 & 1.37 & 0.68 & 0.30 & 1.04\\
		\cline{2-8}
		& \multirow{2}{*}{1000} & 0.5&0.228 & 1.42 & 0.61 & 0.34 & 0.95\\
		\cline{3-8}
		& & 0.95&0.551 & 1.36 & 0.67 & 0.23 & 1.09\\
		\cline{1-8}
		\multirow{4}{*}{1} & \multirow{2}{*}{300} & 0.5&0.249 & 1.16 & 0.85 & 0.76 & 0.95\\
		\cline{3-8}
		& & 0.95&0.312 & 1.19 & 0.90 & 1.02 & 0.92\\
		\cline{2-8}
		& \multirow{2}{*}{1000} & 0.5&0.215 & 1.40 & 0.75 & 0.61 & 0.95\\
		\cline{3-8}
		& & 0.95&0.289 & 1.27 & 0.84 & 0.76 & 0.95\\
		\hline
	\end{tabular}
	\caption{$\hat{\MISE}$ of D-vine quantile regression and the relative performances of the other estimation methods based on data generated by Scenario \textbf{M5} with different parameter settings.}
	\label{tab:M5Comp}
\end{table}
%\begin{table}[ht]
%	\centering
%	\small
%	\begin{tabular}{|c|c|c||}
%		\hline
%		$\sigma$ & $n_{train}$ & $\alpha$\\[0.5pt]
%		\hline
%		\multirow{4}{*}{0.1} & \multirow{2}{*}{300} & 0.5\\[0.37pt]
%		\cline{3-3}
%		& & 0.95\\[0.37pt]
%		\cline{2-3}
%		& \multirow{2}{*}{1000} & 0.5\\[0.37pt]
%		\cline{3-3}
%		& & 0.95\\[0.37pt]
%		\cline{1-3}
%		\multirow{4}{*}{1} & \multirow{2}{*}{300} & 0.5\\[0.37pt]
%		\cline{3-3}
%		& & 0.95\\[0.37pt]
%		\cline{2-3}
%		& \multirow{2}{*}{1000} & 0.5\\[0.37pt]
%		\cline{3-3}
%		& & 0.95\\[0.37pt]
%		\hline
%	\end{tabular}
%	\hspace{-2.8mm}
%	\begin{tabular}{r|r|r|r|}
%		\hline
%		$\hat{\MISE}_{DVQR}$ & $\hat{\RMISE}_{LQR}$ & $\hat{\RMISE}_{NPQR}$ & $\hat{\RMISE}_{SPQR}$ \\ 
%		\hline
%  0.209 & 1.39 & 0.41 & 0.95 \\ 
%  \hline
%  0.514 & 1.37 & 0.30 & 1.04 \\ 
%  \hline
%  0.228 & 1.42 & 0.34 & 0.95 \\ 
%  \hline
%  0.551 & 1.36 & 0.23 & 1.09 \\ 
%  \hline
%  0.249 & 1.16 & 0.76 & 0.95 \\ 
%  \hline
%  0.312 & 1.19 & 1.02 & 0.92 \\ 
%  \hline
%  0.215 & 1.40 & 0.61 & 0.95 \\ 
%  \hline
%  0.289 & 1.27 & 0.76 & 0.95 \\
%		\hline
%	\end{tabular}
%	\caption{Empirical $\hat{\MISE}$ of D-vine quantile regression and the relative performances of the other estimation methods based on data generated by Scenario \textbf{M5} with different parameter settings.}
%	\label{tab:M5Comp}
%\end{table}

\begin{figure}[htbp]
	\centering
	\includegraphics[width=0.45\textwidth]{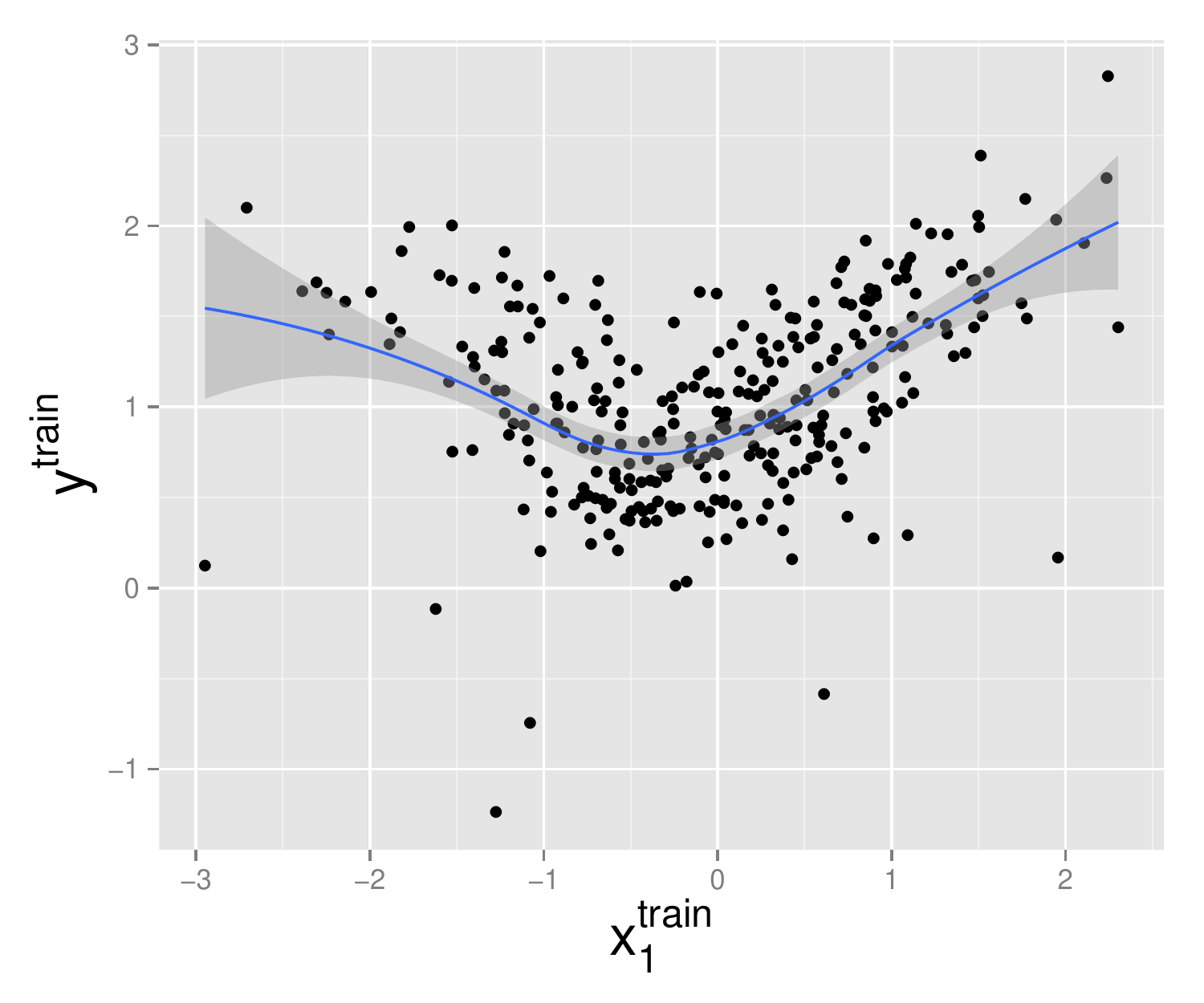}
	\includegraphics[trim=0cm 0.7cm 0cm 2.0cm,clip,width=0.45\textwidth]{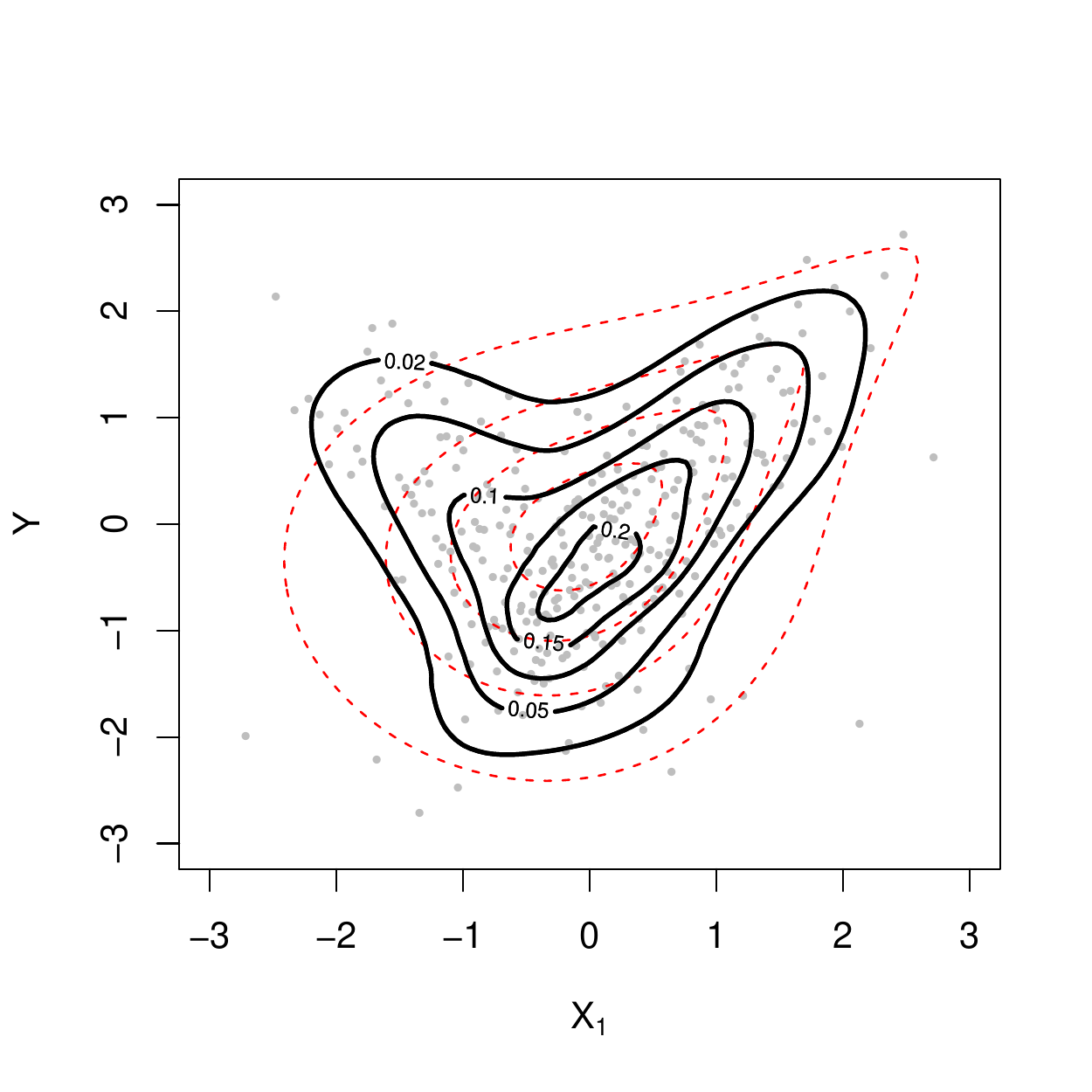}
	\caption{Left: Exemplary scatterplot between $Y$ and $X_1$ in the setting $\sigma=0.1$ and $n_{train}=300$ of Scenario \textbf{M5}, together with a locally weighted scatterplot smoothing line. Right: Empirical nonparametric contour plot of the copula between $X_1$ and $Y$ with standard normal margins (solid) and the corresponding contour plot of the fitted copula (dashed).}
	\label{fig:M5_cop}
\end{figure}

Motivated by \cite{dette2014some} who argue that a non-monotonic relationship between the response and the predictors cannot be modeled by a parametric copula, we consider Scenario \textbf{M5} as a case where no parametric D-vine (or R-vine) is able to perfectly capture the model's dependence structure, such that the model is misspecified. This becomes noticeable when looking at the estimated mean squared error of D-vine quantile regression which is clearly larger than in the previous examples. Also, we observe that the order of magnitude of the errors does not change when moving from a small to a large sample size implying a model bias. This model bias also seems to be visible when linear and semiparametric quantile regressions are used. All three models are quite close in performance with the linear model falling a bit behind. However, we observe that in this scenario nonparametric and boosting additive quantile regression methods excel. Imposing no restrictions on the model assumptions, NPQR generates relative errors ranging between 0.23 and 1.02. This motivates further research to facilitate the inclusion of nonparametric pair-copulas \citep{nagler2015evading} in the construction of the D-vine used for quantile regression to accommodate non-monotonicities. The need for this is also underlined in the setting $\sigma=0.1$ and $n_{train}=300$ by an exemplary scatterplot between $Y$ and $X_1$ (which is the first variable chosen by the D-vine regression algorithm), displayed in the left panel of \autoref{fig:M5_cop} (together with a locally weighted scatterplot smoothing line).

We clearly see a non-monotonic relationship between $X_1$ and $Y$. Currently, no pair-copula implemented in the \texttt{VineCopula} package is able to model such a dependence. The solid lines of the right panel of \autoref{fig:M5_cop} display the empirical nonparametric contour plot of the corresponding copula with standard normal margins generated by the function \texttt{kdecop} of the package \texttt{kdecopula} \citep{kdecopula}. In dashed lines the corresponding contour plot of the fitted parametric copula is shown (Joe copula with $\tau\approx0.25$). The unsatisfying model fit is obvious and explains the rather high estimated MISE values. With this in mind, it is also understandable that a larger training sample size does not help to improve the model fit and prediction accuracy of D-vine quantile regression for this example. However, we observe that the nonparametrically estimated copula manages to model the non-monotonic dependence of the data quite well (for visualization purposes we added the data points transformed to have standard normal margins as well). Hence, by using a nonparametric copula to model the dependence of the pair $(Y,X_1)$, a model misspecification as described in \cite{dette2014some} would be avoided. Further, as discussed in \cite{nagler2015evading} in detail, by modeling only bivariate copulas nonparametrically the dreaded curse of dimensionality is evaded.

\section{Application to the CDS data set}\label{sec:cdsapp}

As an application of D-vine quantile regression to real data we want to exploit interdependencies in the financial market in order to set up models that are able to forecast performances of single institutions as well as construct global stress tests. For these purposes we consider a data set containing 1371 daily observations (01/04/2006 -- 10/25/2011) of log-returns of credit default swap (CDS) spreads with 5 year maturity of 38 European, US American and Asian-Pacific financial institutions in the banking and insurance sectors. This data set has already been analyzed by \cite{brechmann2013conditional}, who argue that CDS spreads are a viable and accurate measure of a company's creditworthiness. After applying an appropriate GARCH model to each of the univariate time series in order to get approximately i.i.d.\ residuals, \cite{brechmann2013conditional} perform stress tests. By sampling from a conditional C-vine the authors stress one company at a time (i.e.\ setting it to its 90\%/95\%/99\% quantile) and examine the impact on the other institutions conditioned on this stress event. With our method we can go even further and consider scenarios where more than one company is in distress. This allows us to investigate the spillover effects of a financial crisis in a certain region or branch to other regions and branches. Additional to the stress tests we use D-vine quantile regression models later in this chapter in order to predict an institution's CDS spread log-returns given its past values and the log-returns of peer companies. In an out-of-sample test we demonstrate the competitiveness of our approach once again.

\subsection{Stress testing}

\begin{figure}[tb]
	\centering
	\includegraphics[trim=0.8cm 0.8cm 0.4cm 0.6cm,clip,width=0.24\textwidth]{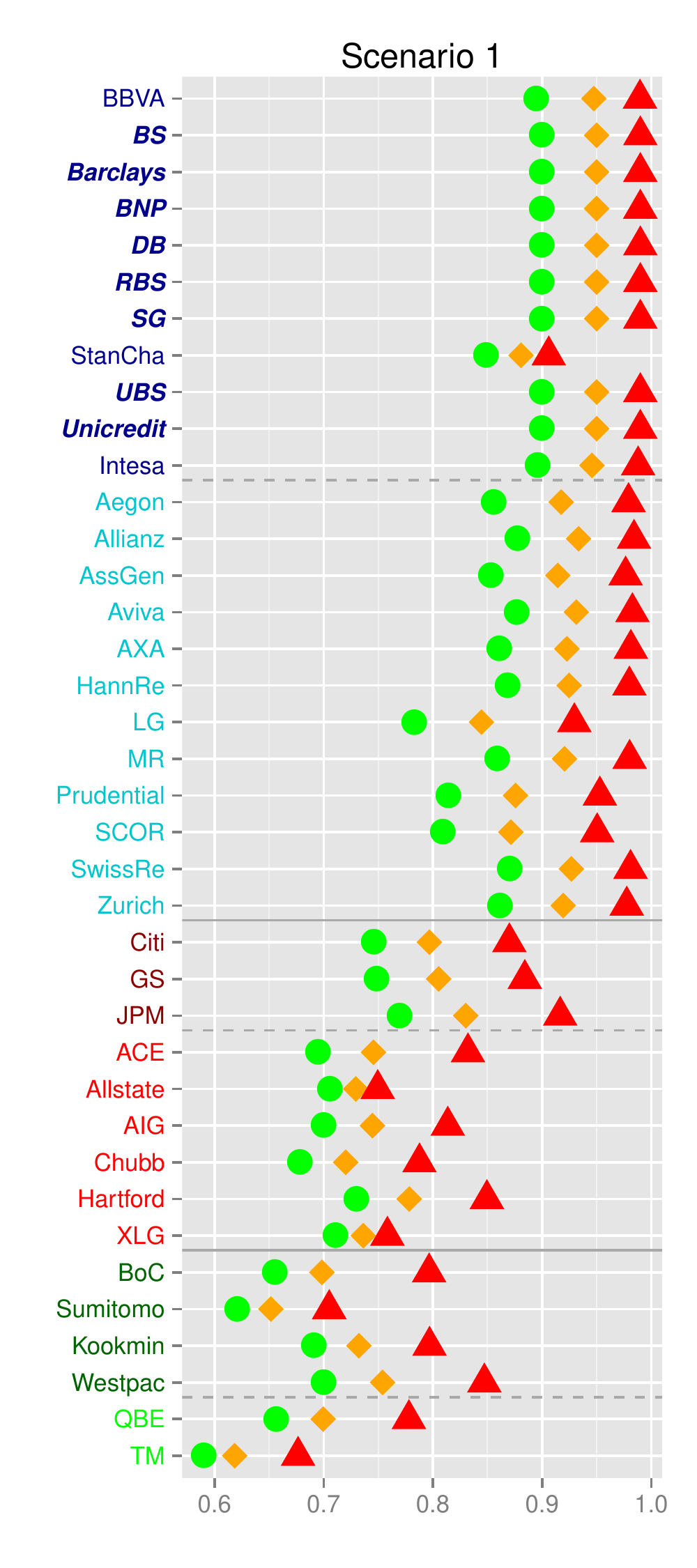}
	\includegraphics[trim=0.8cm 0.8cm 0.4cm 0.6cm,clip,width=0.24\textwidth]{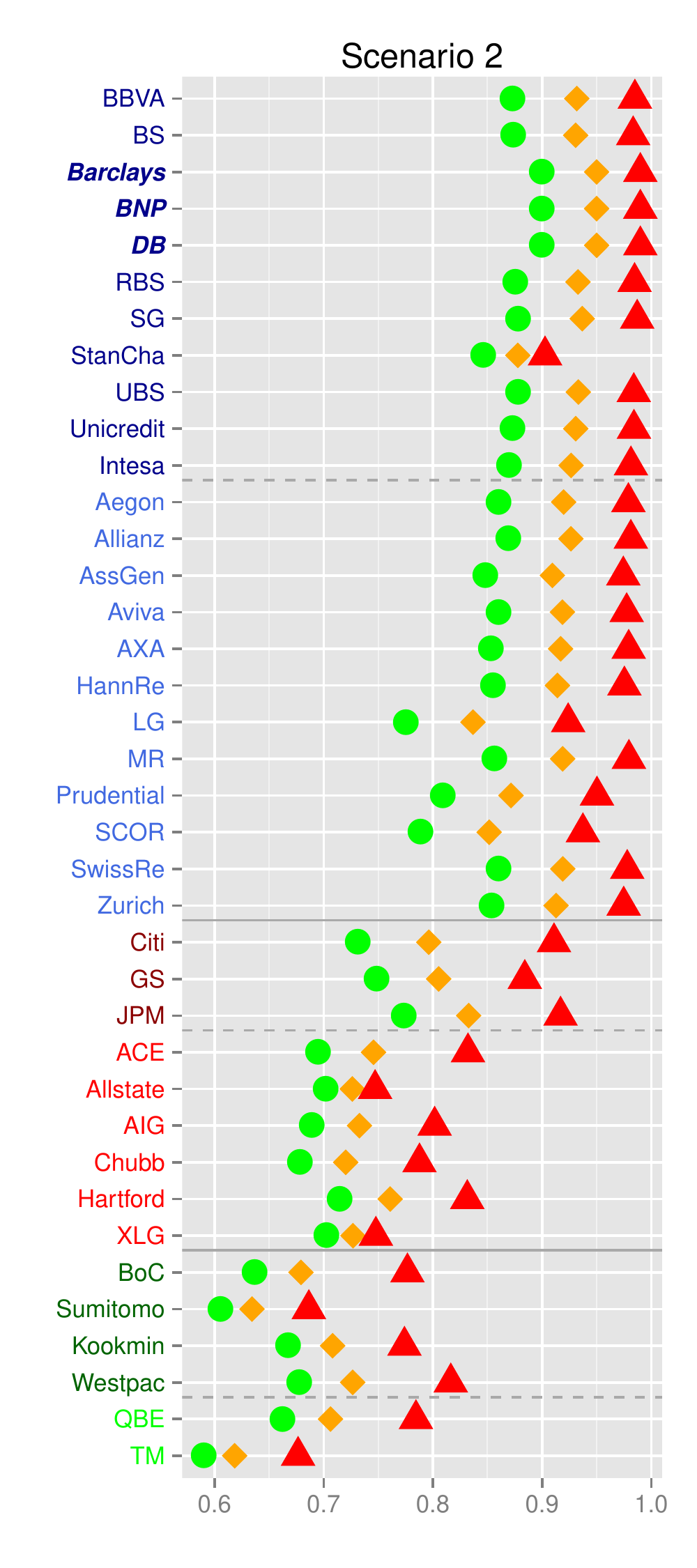}
	\includegraphics[trim=0.8cm 0.8cm 0.4cm 0.6cm,clip,width=0.24\textwidth]{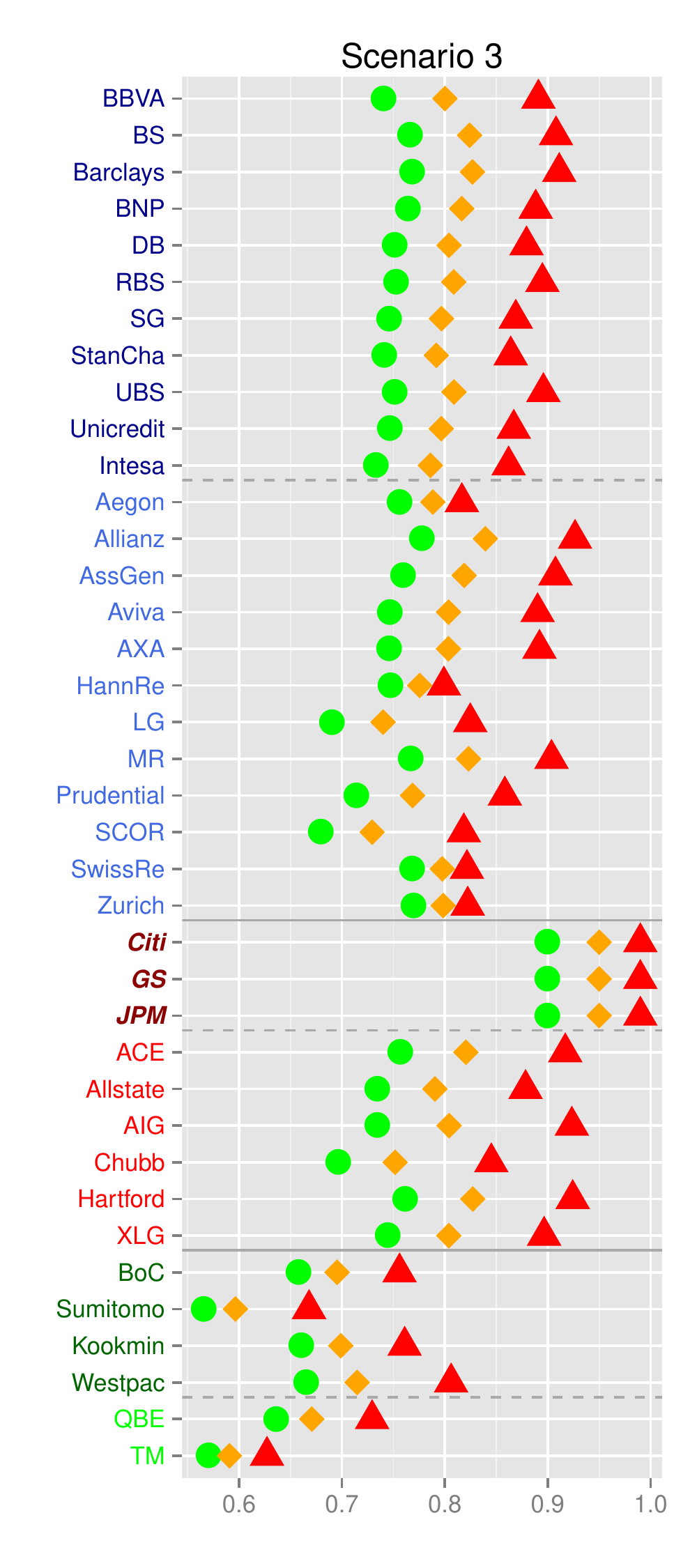}
	\includegraphics[trim=0.73cm 0.6cm 0.15cm 0.18cm,clip,width=0.242\textwidth]{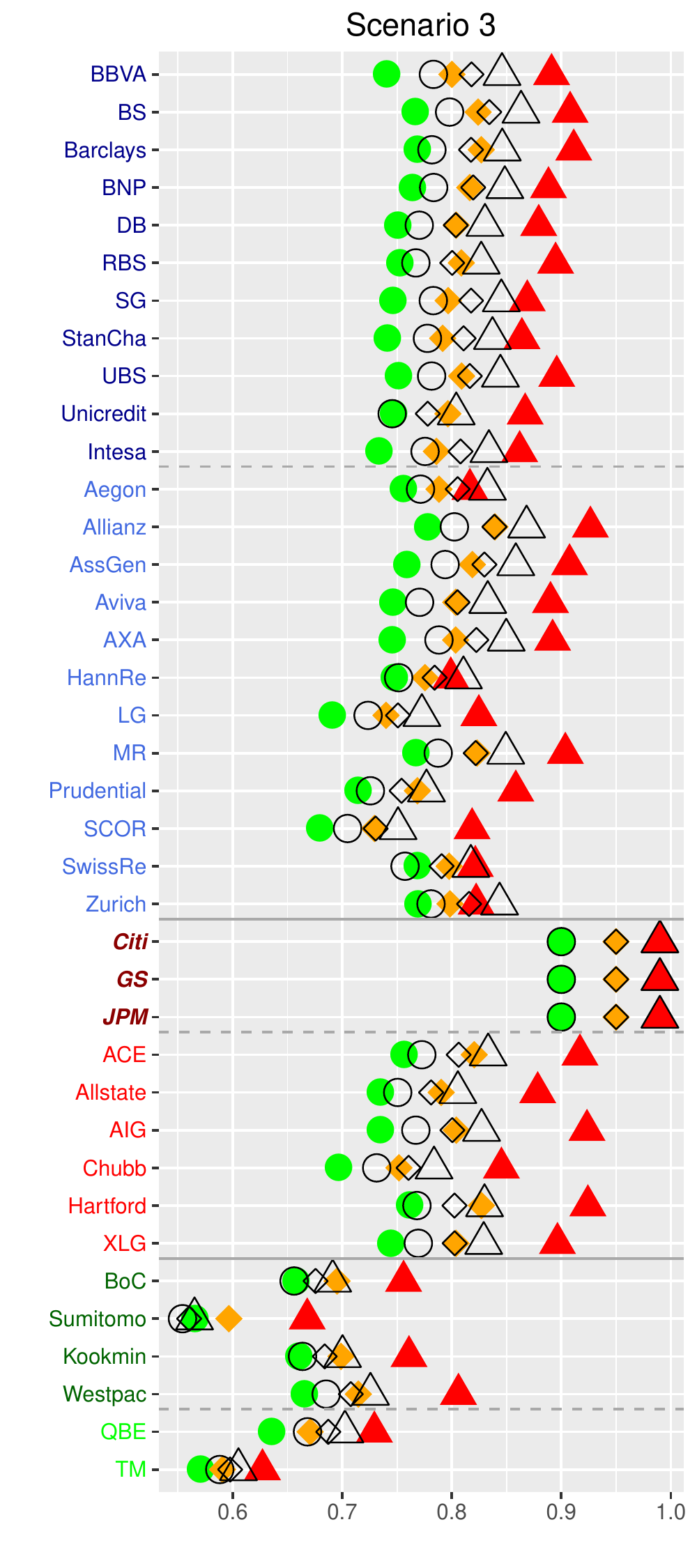}
	\caption{Stress tests stressing European systemic banks (left), major European banks (middle) and US banks (right). For each institution the predicted median values for the three stress levels are represented by circles (moderate stress with $\kappa=0.9$), diamonds (severe stress with $\kappa=0.95$) and triangles (extreme stress with $\kappa=0.99$).}
	\label{fig:StressTest}
\end{figure}

The financial institutions considered in the stress test are 18 banks and 20 (re-)insurers from the regions USA, Europe and Asia-Pacific:
\begin{itemize}
	\item \textbf{Banks}: \textit{USA} (Goldman Sachs (GS), JP Morgan Chase (JPM), Citigroup), \textit{Europe} (Banco Bilbao Vizcaya Argentaria (BBVA), Banco Santander (BS), Barclays, BNP Paribas, Deutsche Bank (DB), Intesa Sanpaolo, Royal Bank of Scotland (RBS), Soci\'et\'e G\'en\'erale	(SG), Standard Chartered (StanCha), UBS, Unicredit), \textit{Asia-Pacific} (Bank of China (BoC), Kookmin Bank, Sumitomo Mitsui, Westpac Banking)
	\item \textbf{(Re-)Insurers}: \textit{USA} (ACE, Allstate, American International Group (AIG), Chubb, Hartford Financial Services, XL Group), \textit{Europe} (Aegon, Allianz, Assicurazioni Generali, Aviva, AXA, Hannover R\"uck (HR), Legal \& General (LG), Munich Re (MR), Prudential, SCOR, Swiss Re (SR), Zurich Insurance), \textit{Asia-Pacific} (Tokio Marine (TM),QBE Insurance)
\end{itemize}
%\begin{itemize}
%	\item \textbf{Banks}:
%	\begin{itemize}
%		\item \textit{USA}: Citigroup, Goldman Sachs (GS), JP Morgan Chase (JPM)
%		\item \textit{Europe}: Banco Bilbao Vizcaya Argentaria (BBVA), Banco Santander (BS), Barclays,
%		BNP Paribas, Deutsche Bank (DB), Intesa Sanpaolo, Royal Bank of Scotland (RBS), Soci\'et\'e G\'en\'erale
%		(SG), Standard Chartered (StanCha), UBS, Unicredit
%		\item Asia-Pacific: Bank of China (BoC), Kookmin Bank, Sumitomo Mitsui, Westpac Banking
%	\end{itemize}
%	\item \textbf{(Re-)Insurers}:
%	\begin{itemize}
%		\item \textit{USA}: ACE, Allstate, American International Group (AIG), Chubb, Hartford Financial Services, XL Group
%		\item \textit{Europe}: Aegon, Allianz, Assicurazioni Generali, Aviva, AXA, Hannover R\"uck (HR),
%		Legal \& General (LG), Munich Re (MR), Prudential, SCOR, Swiss Re (SR), Zurich
%		Insurance
%		\item \textit{Asia-Pacific}: QBE Insurance, Tokio Marine (TM)
%	\end{itemize}
%\end{itemize}
We consider three stress scenarios, corresponding to crises originating in different sectors, and investigate the resulting spillover effects. For this, we proceed as in \cite{brechmann2013conditional} and remove serial dependencies from each of the 38 univariate time series by fitting adequate GARCH models. The resulting residuals $r_j$, $j=1,\ldots,38$, which are approximately independent and distributed according to their model's estimated innovations distribution $\hat F_j$, carry the information about the dependence structure between the institutions. We consider company $j$ to be stressed at level $\kappa\in(0,1)$, if its residual $r_j$ takes on the $100\kappa\%$-quantile of its innovation distribution $\hat F_j$, i.e.\ $r_j=\hat F^{-1}_j(\kappa)$. This is equivalent to the PIT transformed variable $u_j:=\hat F_j(r_j)$ taking on value $\kappa$. Likewise, we are interested in the resulting predicted quantile levels of the non-stressed companies. This allows us to directly work on the u-scale and consider the PIT transformed variables $u_j:=\hat F_j(r_j)$, $j=1,\ldots,38$, and their dependencies.\\
In Scenario 1, we analyze the effect of stressing the European \textit{systemic} banks as specified by \cite{risk2009global} at different stress levels. Therefore, we stress the banks Banco Santander, Barclays,
BNP Paribas, Deutsche Bank, Royal Bank of Scotland,  Soci\'et\'e G\'en\'erale, UBS and Unicredit at level $\kappa\in\{0.9,0.95,0.99\}$ (corresponding to moderate, severe and extreme stress scenarios) and use D-vine quantile regression to estimate the conditional medians of the remaining institutions conditioned on this stress event. This way, we can assess the spillover effect to other sectors and regions. The left panel of \autoref{fig:StressTest} shows the results of the stress test of Scenario 1. For each institution the predicted median values for the three stress levels are coded by circles (moderate stress with $\kappa=0.9$), diamonds (severe stress with $\kappa=0.95$) and triangles (extreme stress with $\kappa=0.99$). For visualization, the currently stressed institutions' names are printed in \textit{\textbf{bold and italic}}. Further, solid lines separate the geographical regions (Europe in the upper, USA in the middle and Asia-Pacific in the lower panel), while dashed lines separate banks (upper) from insurance companies (lower).\\
%For visualization, the names of the institutions are colored according to their groups ({\color{blue}Europe blue}, {\color{red}USA red}, {\color{green}Asia-Pacific green}, banks dark colors, insurances light colors) and the currently stressed institutions' names are printed in \textit{\textbf{bold and italic}}.
We observe that the spillover effect is strongest for European insurances with predicted median values of up to 0.98 (Allianz and Aviva) for the extreme stress scenario. The comparably small values of the British bank Standard Chartered are explained by the fact that according to their annual report of 2014 (\url{https://www.sc.com/annual-report/2014/documents/SCB_ARA_2014_full_report.pdf}) 90\% of the bank's income and profits are earned in Asia, Africa and the Middle East. Similar arguments holds for the British insurance company L\&G with operations in Asia and the United States (\url{http://www.legalandgeneralgroup.com/all-our-sites/}). Another group that is affected quite strongly by this stress scenario are the US banks with predicted median values exceeding 0.9 in the extreme stress case. However, we observe that the geographic spillover effect is stronger than the institutional one, because the effect on European insurance companies is stronger than the effect on US banks. US insurance companies as well as the Asian-Pacific market are also affected by the stress scenario, but not as severe as the other groups.\\
It is interesting to see that in Scenario 2, where we only stress the three major European banks Barclays, BNP Paribas and Deutsche Bank, the results of the stress test are very similar to those of Scenario 1. We can conclude, that for a crisis to evolve it suffices that only few but important banks default. %paragraph?
In the last scenario, we analyze the spillover effect of a default of the US American banking system (see third panel of \autoref{fig:StressTest}). Therefore, we stress the banks Citigroup, Goldman Sachs and JP Morgan Chase at level $\kappa\in\{0.9,0.95,0.99\}$ and estimate the conditional medians of the remaining institutions conditioned on this stress event. Again, we see the quite strong interconnectedness between US banks and insurance companies, as well as between US banks and the European market, and observe rather weak spillover effects on the Asian-Pacific sector.\\
Finally, in the right panel of \autoref{fig:StressTest}, we compare the stress testing results of D-vine quantile regression to those of linear quantile regression for Scenario 3. Additional to the filled symbols indicating the results of DVQR, we also added the predicted medians of LQR with empty symbols, where circles again denote moderate stress ($\kappa=0.9$), diamonds severe stress ($\kappa=0.95$) and triangles extreme stress ($\kappa=0.99$). We see that for almost all companies linear regression overestimates the moderate stress results and underestimates the extreme ones. This reflects the fact that Gaussian dependence structure implied by linear quantile regression fails to imitate the tail dependence that is typically exhibited by financial data such as these CDS log returns. Another flaw of linear quantile regression observable from the plot is that due to the linearity of the model the median predictions for the three different stress levels seem to always have a similar distance to each other. The predictions of the D-vine based quantile regression appear as much more flexible with some narrower, wider and skewed sets of predictions for the three stress levels. The linear quantile regression results of Scenarios 1 and 2 allow for similar conclusions and are therefore omitted here.

\subsection{Out-of-sample quantile prediction}

In order to assess the prediction performance of D-vine quantile regression we conduct an out-of-sample test. For this, we consider the raw CDS log-return data (i.e.\ no time series models were fitted). The goal is to predict the companies' conditional Value-at-Risk at several time points, i.e.\ the quantiles of $Y_t$ (the CDS log-returns of company $Y$ at time $t$) given its past values $Y_{t-1},\ldots,Y_{t-l}$ (for some lag $l\geq1$) and the past and current values of its peer companies $\Xb_t,\ldots,\Xb_{t-l}$. In the out-of-sample test, we split the data set at a certain time point into a training data set, based on which we will fit the different quantile regression models, and an evaluation data set used assess the performance of the methods. As a split point we choose the date November 8th, 2010, such that the training data set contains $n_{train}=1121$ observations and the evaluation data set contains $n_{eval}=250$ observations. Note that this sample is not i.i.d. However, since we can assume that it fulfills the strong mixing condition due to low autocorrelations for lags higher than 3 \citep{hansen2008uniform}, the approach of estimating the marginals nonparametrically is justified.

One by one, we consider the CDS log-returns of each of the 38 financial institutions as responses. For a fixed response it is not possible to use all of the remaining 37 institutions as predictors since the computational times for the nonparametric and semiparametric quantile regression methods would be too high. Therefore, we reduce the number of predictors by running the D-vine regression algorithm on all of the remaining institutions with lag value $l$ equal to 2 and choosing only the meaningful institutions as potential covariates. This D-vine based screening of the covariates results in a total of 5-10 variables selected, depending on the response institution. No lags higher than $l=1$ were chosen.

Using the selected covariates, for each of the 38 institutions we fit the five previously introduced quantile regression methods on the training data and afterwards apply them to the out-of-sample data set to estimate conditional quantiles at levels $\alpha\in\{0.5,0.01,0.99,0.995\}$. Note that for $\alpha=0.995$ the predicted quantile corresponds to the conditional Value-at-Risk at the $99.5\%$-level. 

In this out-of-sample test the true regression quantiles are unknown, we just observe a realization for each day. A possible way of evaluating the predicted $\alpha$-quantiles for some $\alpha\in(0,1)$ is given by the averaged tick-loss $L^j_{\alpha,m}$ \cite[see e.g.][]{komunjer2013quantile}, which, for company $j$, evaluation sample size $n_{eval}$, observations $y^{(i)}$ and quantiles $q_{\alpha,m}^{(i)}$ predicted by using method $m$, $i=1,\ldots,n_{eval}$, is defined by
\[
L^j_{\alpha,m}=\frac{1}{n_{eval}}\sum_{i=1}^{n_{eval}}\rho_{\alpha}(y^{(i)}-\hat q_{\alpha,m}^{(i)}),
\]

where again $\rho_{\alpha}(y)=y(\alpha-\mathbbm{1}(y<0))$ denotes the check or \textit{tick} function.

\autoref{tab:OOS_tickloss} displays the tick-loss $L_{\alpha,m}$ of predicted quantiles at levels $\alpha\in\{0.5,0.01,0.99,0.995\}$ for the prediction methods $m\in\{\text{DVQR}, \text{LQR}, \text{BAQR}, \text{NPQR}, \text{SPQR}\}$, averaged over the 38 financial institutions, i.e.\ $L_{\alpha,m}=1/38\sum_{j=1}^{38}L^j_{\alpha,m}$.

\begin{table}[ht]
	\centering
	\begin{tabular}{|lccccc|}
		\hline
		quantile & DVQR & LQR & BAQR & NPQR & SPQR \\ 
		\hline
		0.5 & 6.14 & 6.12 & 6.15 & 7.68 & 6.14 \\ 
		0.01 & 0.63 & 0.99 & 0.69 & 0.75 & 0.61 \\ 
		0.99 & 0.72 & 1.11 & 0.82 & 0.87 & 0.71 \\ 
		0.995 & 0.43 & 0.73 & 0.50 & 0.55 & 0.43 \\ 
		\hline
	\end{tabular}
	\caption{Results of the out-of-sample test predicting the CDS log-returns: tick-loss $L_{\alpha,m}$ of predicted quantiles at levels $\alpha\in\{0.5,0.01,0.99,0.995\}$ for different prediction methods averaged over the 38 financial institutions.}
	\label{tab:OOS_tickloss}
\end{table}

A vast majority of the pair-copulas selected by the algorithm for the estimation of the D-vines were t copulas (which is not surprising for a financial data set). As we have seen in \autoref{tab:t5Comp}, linear quantile regression works quite well for predicting the median of such models. This might explain why it manages to slightly outperform all other methods in the conditional 0.5-quantile prediction. Moving away from the median to the more extreme quantiles its lack to describe the fat tails of the t copula become apparent, leaving linear quantile regression far behind all the other methods. Further we can observe, that for this data set the D-vine and semiparametric quantile regression method perform very similar and yield better results than their competitors for the prediction of extreme quantiles. The minimal advantage of SPQR might be explained by the fact that the objective of SPQR is exactly to minimize the above loss function. Boosted additive quantile regression succeeds in improving the linear approach for the quantiles in the tails yielding the third-best results. With as many predictors as 5-10, the nonparametric method already seems to run into the curse of dimensionality and thus produces the worst result for the median prediction and the second worst for the prediction of the quantiles in the tails.

All in all, this real data application has supported the findings of the simulation study of \autoref{ch:ss}, that D-vine quantile regression is a reasonable tool for estimating conditional quantiles.

\section{Conclusions and further research}\label{sec:conclusion}

A new method to predict conditional quantiles is proposed. We have seen that the usage of the flexible D-vine class facilitates fast and accurate estimation. Analyzing the log returns of the CDS spreads of international banks and insurances we extend the analysis of \cite{brechmann2013conditional}. While they analyze the spillover effects stressing only one institution by simulating from a conditioned C-vine, with our new method we are able to perform stress tests that are conditioned on multiple banks and insurances being in distress. In our analysis we found out that the spillover effect is mainly driven by geography, so that European banks have a greater influence on European insurances than on US American banks. Further, the claim of \cite{brechmann2013conditional} that US banks have a stronger influence on the international financial market than European banks is not supported by our analysis. Stressing the major European banks had a greater overall impact on the financial system than stressing the US American banks.\\
Moreover, we have seen that there is still room for future research. With the bivariate pair-copulas implemented in the package \texttt{VineCopula} there is currently no possibility to model non-monotonic dependencies between a response and its predictors as is already pointed out in \cite{dette2014some}. A remedy for the resulting issue of possible misspecification is the inclusion of nonparametric pair-copulas in the construction of the D-vine used for quantile regression. A further topic of ongoing research is the implementation of D-vines with mixed discrete and continuous margins as discussed in \cite{stober2013simplified} in order to yield quantile regression models allowing for discrete as well as continuous responses and predictors.

\section*{Appendix}
\begin{appendix}

\section{D-vine regression algorithm}
\label{ch:Algo}
%\begin{alg}[D-vine regression algorithm]\mbox{}\\ 
\textbf{Input:} pseudo copula data $\vb=(v^{(i)})_{i=1,\ldots,n}$, $\Uc=(u^{(i)}_{j})_{j=1,\ldots,d,\ i=1,\ldots,n}$
\begin{algorithmic}[1]
  \STATEx Initialize global variables:
  \STATE $I_0 \leftarrow \left\{1,\ldots,d\right\}$
  \STATE $\verb+global.max.cll+ \leftarrow -\infty$
  \STATEx \mbox{}
	\STATEx Step 1: 
	\FOR{$j\in I_0$}
	  \STATE Use \texttt{BiCopSelect} % (\autoref{alg:BiCopSelect})
	   to estimate $C_{V,U_j}$, the AIC-optimal pair-copula for $(V,U_j)'$ based on the data $(\vb,\ub_j)$
	  \STATE $\cll_j \leftarrow \cll(\hat c_{V,U_j};\vb,\ub_j)$ 
	\ENDFOR
	\STATE $l_1 \leftarrow \argmax_{j\in I_0}\cll_j$%\texttt{ \# First variable of the D-vine}
	\STATEx Update global variables:
	\STATE $\verb+global.max.cll+ \leftarrow \max_{j\in I_0}\cll_j$
	\STATE $I_1 \leftarrow I_0\backslash l_1$
	\STATEx \mbox{}
	\STATEx Step $k$: (the first tree of the current D-vine with $k-1$ covariates has the order $V$--$U_{l_1}$--$\ldots$--$U_{l_{k-1}}$ (see the black graph in \autoref{fig:Algo_Step_k}). $I_{k-1}$ contains the indices of the covariates that have not yet been chosen for the model and \verb+global.max.cll+ is equal to the conditional log-likelihood of the current model)
	\FOR{$j\in I_{k-1}$}
	  \STATE Estimate $C_{V,U_j|U_{l_1},\ldots,U_{l_{k-1}}}$ and $C_{U_{l_i},U_j|U_{l_i+1},\ldots,U_{l_{k-1}}}$, $i=1,\ldots,k-1$, i.e.\ the pair-copulas needed to extend the current D-vine to a D-vine with order $V$--$U_{l_1}$--$\ldots$--$U_{l_{k-1}}$--$U_j$ (compare \autoref{fig:Algo_Step_k})
%	  \STATE Extend the current D-vine by adding $X_i$ as a leaf, so that the resulting first tree has the order $Y$--$X_{l_1}$--$\ldots$--$X_{l_{k-1}}$--$X_i$
	  \STATE $\cll_j \leftarrow \cll(\hat c_{V,U_{l_1},\ldots,U_{l_{k-1}},U_j};\vb,\ub_{l_1},\ldots,\ub_{l_{k-1}},\ub_j)$
	\ENDFOR
	\IF{$\max_{j\in I_{k-1}}\cll_j<\tt{global.max.cll}$ } 
	  \STATE \textbf{return} D-vine with order $V$--$U_{l_1}$--$\ldots$--$U_{l_{k-1}}$ 
	\ENDIF
	\STATE $\verb+global.max.cll+ \leftarrow \max_{j\in I_0}\cll_j$
	\STATE $l_k \leftarrow \argmax_{j\in I_{k-1}}\cll_j$
	\STATE $I_k \leftarrow I_{k-1}\backslash l_k$
  \STATEx \mbox{}
  \STATE After step $d$ has been exercised, \textbf{return} D-vine with order $V$--$U_{l_1}$--$\ldots$--$U_{l_d}$
	\end{algorithmic}
%\label{alg:D_Vine_seq}
%\end{alg}

In lines 5 and 12 of the algorithm, the AIC- and BIC-corrected conditional log-likelihoods may be used instead of the regular $\cll$. In this case all maximum functions have to be replaced by minimum functions.\\
%The algorithm repeatedly uses the function \texttt{BiCopSelect} of the package \texttt{VineCopula}, which fits the $\BIC$-optimal pair-copula to copula data after testing for independence (for details see \cite{genest2007everything}):
%
%\begin{alg}[\texttt{BiCopSelect}]\mbox{}\\ 
%\textbf{Input:} i.i.d. copula data $(u_{1}^{(i)},u_{2}^{(i)})$, $i=1,\ldots,n$; significance level for independence test $\beta$.\\
%\begin{algorithmic}[1]
%  \STATE Calculate the empirical Kendall's Tau $\hat{\tau}$ and the test statistic $t=\sqrt{\frac{9n(n-1)}{2(2n+5)}}|\hat{\tau}|$.
%	\IF{$2(1-\Phi(t))\geq\beta$}
%	\STATE \textbf{Return} the independence copula $c_{\pi}$.
%	\ENDIF
%  \FOR{each pair-copula family}
%	\STATE $\theta_{ML} = \argmax_{\theta\in\Theta}\prod_{i=1}^n{c(u_1^{(i)},u_2^{(i)};\theta)}$.
%	\STATE $\BIC=-2\sum_{i=1}^n{\ln(c(u_1^{(i)},u_2^{(i)};\theta_{ML}))}+\ln(n)|\theta_{ML}|$.
%	\ENDFOR
%  \STATE \textbf{Return} the copula with the lowest $\BIC$-value with its MLE parameter $\theta_{ML}$.
%	\end{algorithmic}
%\label{alg:BiCopSelect}
%\end{alg}
%Here, $|\theta_{ML}|$ denotes the number of parameters of the copula family under consideration.\\
Regarding the possible pair-copula families to be selected in line 4 of the code, the families currently implemented in the \texttt{VineCopula} package are Gaussian, t*, Clayton, Gumbel, Frank, Joe, BB1*, BB6*, BB7*, BB8* and Tawn* with respective rotations. Families marked with a star are two-parametric, while the others are one-parametric.\\
\section{Analytically derived conditional copula quantiles}\label{sec:AnalyticalCondQuant}

\subsection{Conditional copula quantile function based on the Gaussian copula}
It is a well known fact that the conditional distribution of a multivariate normal distributed random vector again is normally distributed with shifted mean and covariance matrix \citep{brachinger1996multivariate}. Therefore, assuming 
\begin{equation}\label{eq:Gauss_spec}
	(Y,X_1,\ldots,X_d)\sim N\left((\mu_Y,\mathbf{\mu_\Xb})',\begin{pmatrix}
		\sigma^2_Y & \Sigma_{Y,\Xb}  \\ 
		\Sigma_{Y,\Xb}' & \Sigma_{\Xb,\Xb}  \\ 
	\end{pmatrix}\right),
\end{equation}
the conditional quantile of $Y$ at level $\alpha\in(0,1)$ given $\Xb=\xb$ can be calculated as
\[
F^{-1}_{Y|\Xb}(\alpha|\xb)=\Phi^{-1}(\alpha;\mu_c(\xb),\sigma^2_c),
\]
where $\Phi(\cdot;\mu,\sigma^2)$ denotes the distribution function of a $N(\mu,\sigma^2)$-distributed random variable and the conditional parameters $\mu_c(\xb)$ and $\sigma^2_c$ are given as
\[
\mu_c(\xb):=\mu_Y+\Sigma_{Y,\Xb}\Sigma_{\Xb,\Xb}^{-1}(\xb-\mathbf{\mu_\Xb}),
\]
and
\[
\Sigma_c:=\sigma_Y^2-\Sigma_{Y,\Xb}\Sigma_{\Xb,\Xb}^{-1}\Sigma_{Y,\Xb}'.
\]
Hence, by inverting \autoref{eq:cond_quantile} the conditional quantile of the Gaussian copula is given as
\[
C^{-1}_{V|U_{1},\ldots, U_{d}}(\alpha|u_{1},\ldots,u_{d})=\Phi(\Phi^{-1}(\alpha;\mu_c(\xb),\sigma^2_c);\mu_Y,\sigma_Y^2),
\]
where $\xb=\left(\Phi^{-1}(u_1;\mu_1,\sigma_1^2),\ldots,\Phi^{-1}(u_d;\mu_d,\sigma_d^2)\right)'$.

\subsection{Conditional copula quantile function based on the Student's t copula}
Similarly, following \cite{kotz2004multivariate}, we know that the conditional distribution of a multivariate t distributed random vector with $\nu$ degrees of freedom again is t distributed with shifted mean, covariance matrix and degrees of freedom. Therefore, with the same partitions of the mean and covariance matrix as in \autoref{eq:Gauss_spec}, the conditional quantile of $Y$ at level $\alpha\in(0,1)$ given $\Xb=\xb$ can be calculated as
\[
F^{-1}_{Y|\Xb}(\alpha|\xb)=t^{-1}(\alpha;\nu_c,\mu_c(\xb),\sigma^2_c(\xb)),
\]
where $t(\cdot;\nu,\mu,\sigma^2)$ denotes the distribution function of a $t(\nu,\mu,\sigma^2)$-distributed random variable and the conditional parameters $\nu_c$, $\mu_c(\xb)$ and $\sigma^2_c(\xb)$ are given as
\[
\nu_c:=\nu+d,
\]
\[
\mu_c(\xb):=\mu_Y+\Sigma_{Y,\Xb}\Sigma_{\Xb,\Xb}^{-1}(\xb-\mathbf{\mu_\Xb}),
\]
and
\[
\sigma_c(\xb):=\frac{\nu+(\xb-\mathbf{\mu_\Xb})'\Sigma_{\Xb,\Xb}^{-1}(\xb-\mathbf{\mu_\Xb})}{\nu+d}(\sigma_Y^2-\Sigma_{Y,\Xb}\Sigma_{\Xb,\Xb}^{-1}\Sigma_{Y,\Xb}').
\]
Consequently, the conditional quantile of the t copula is given as
\[
C^{-1}_{V|U_{1},\ldots, U_{d}}(\alpha|u_{1},\ldots,u_{d})=t(t^{-1}(\alpha;\nu_c,\mu_c(\xb),\sigma^2_c(\xb));\nu,\mu_Y,\sigma_Y^2),
\]
where $\xb=\left(t^{-1}(u_1;\nu,\mu_1,\sigma_1^2),\ldots,t^{-1}(u_d;\nu,\mu_d,\sigma_d^2)\right)'$.

\subsection{Conditional copula quantile function based on the two- and three-dimensional Clayton copula}
For the bivariate Clayton copula with parameter $\delta>0$ ($C_{U,V}(u,v)=(u^{-\delta}+v^{-\delta}-1)^{\frac{-1}{{\delta}}}$) the conditional quantile can easily be derived by inverting the h-function, yielding
\[
C_{V|U}^{-1}(\alpha|u)= \left\{ \left( \alpha^{{\frac {-\delta}
		{1+\delta}}}-1\right){u}^{-\delta} +1 \right\} ^{\frac{-1}{\delta}}.
\]
The conditional quantile function of a three-dimensional Clayton copula with parameter $\delta>0$ ($C_{U,V,W}(u,v,w)=(u^{-\delta}+v^{-\delta}+w^{-\delta}-2)^{-\frac{1}{\delta}}$) is given by
\[
C^{-1}_{U|V,W}(\alpha|v,w)=\left\{(\alpha^{-\frac{\delta}{1+2\delta}}-1)(v^{-\delta}+w^{-\delta}-1)+1\right\}^{-\frac{1}{\delta}}.
\]
To derive this, let $(U,V,W)$ follow a three-dimensional Clayton copula with parameter $\delta>0$.
Then,
\[
C_{U,V,W}(u,v,w)=(u^{-\delta}+v^{-\delta}+w^{-\delta}-2)^{-\frac{1}{\delta}}.
\]

\begin{figure}[!htb]
	\centering
	\includegraphics[width=0.7\textwidth]{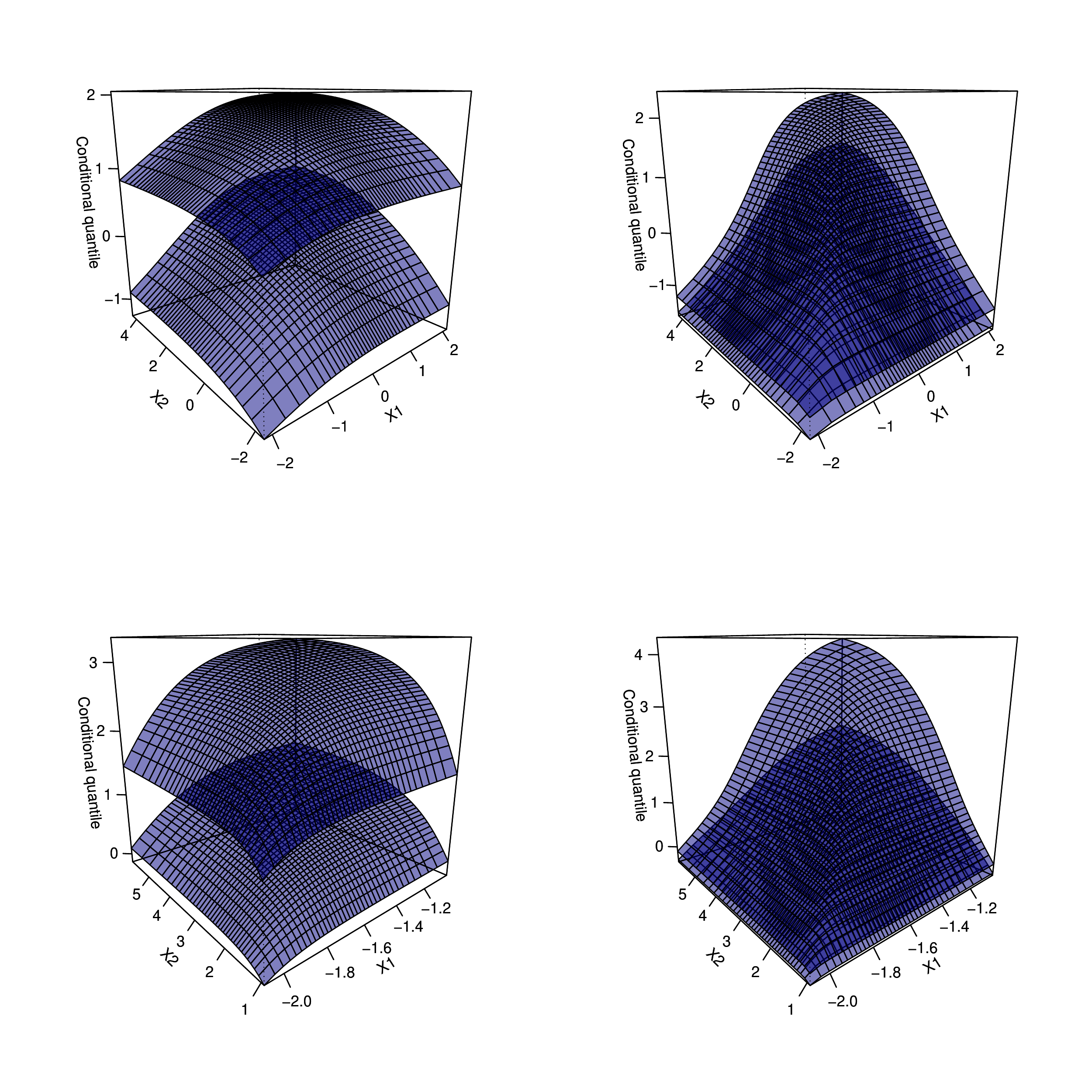}
	\caption{Conditional median and 95\% quantile function of a three-dimensional distribution with underlying Clayton copula with parameter $\delta_1$ (left panels), $\delta_2$ (right panels), and marginals $\mathcal{M}_1$ (upper panels), $\mathcal{M}_2$ (lower panels) as specified in \autoref{tab:scenarios}.}
	\label{fig:Clayton_quantiles}
\end{figure}

In order to derive the conditional distribution of $U$ given $V=v$ and $W=w$ we use a version of \autoref{eq:Joe97}:

\begin{equation}
C_{U|V,W}(u|v,w)=\frac{\partial}{\partial\xi}C_{U,V;W}(C_{U|W}(u|w),\xi;w)\Big|_{\xi=C_{V|W}(v|w)}.
\label{eq:Joe1997}
\end{equation}
For the derivation of $C_{U|W}(u|w)$ we use the fact that $C_{U,W}(u,w)=C_{U,V,W}(u,1,w)=(u^{-\delta}+w^{-\delta}-1)^{-\frac{1}{\delta}}$. Hence,
\begin{align*}
C_{U|W}(u|w)&=\frac{\partial}{\partial w}C_{U,W}(u,w)=\frac{\partial}{\partial w}(u^{-\delta}+w^{-\delta}-1)^{-\frac{1}{\delta}}\\
&=(u^{-\delta}+w^{-\delta}-1)^{-\frac{1}{\delta}-1}w^{-\delta-1},
\end{align*}
and similarly,
\[
C_{V|W}(v|w)=(v^{-\delta}+w^{-\delta}-1)^{-\frac{1}{\delta}-1}w^{-\delta-1}.
\]
The corresponding inverse functions are given by 
\[
C^{-1}_{U|W}(u|w)=\left\{w^{-\delta}\left(u^{-\frac{\delta}{1+\delta}}-1\right)+1\right\}^{-\frac{1}{\delta}}
\]
and
\[
C^{-1}_{V|W}(v|w)=\left\{w^{-\delta}\left(v^{-\frac{\delta}{1+\delta}}-1\right)+1\right\}^{-\frac{1}{\delta}}.
\]
Further,
\[
C_{U,V|W}(u,v|w)=\frac{\partial}{\partial w}C_{U,V,W}(u,v,w)=(u^{-\delta}+v^{-\delta}+w^{-\delta}-2)^{-\frac{1}{\delta}-1}w^{-\delta-1},
\]
and therefore,
\begin{align*}
C_{U,V;W}(u,v;w)&=C_{U,V|W}\left(C^{-1}_{U|W}(u|w),C^{-1}_{V|W}(v|w)|w\right)\\
&=\left\{w^{-\delta}\left(u^{-\frac{\delta}{1+\delta}}-1\right)+1+w^{-\delta}\left(v^{-\frac{\delta}{1+\delta}}-1\right)+1+w^{-\delta}-2\right\}^{-\frac{1}{\delta}-1}w^{-\delta-1}\\
&=\left\{w^{-\delta}\left(u^{-\frac{\delta}{1+\delta}}+v^{-\frac{\delta}{1+\delta}}-1\right)\right\}^{-\frac{1+\delta}{\delta}}w^{-\delta-1}=\left(u^{-\frac{\delta}{1+\delta}}+v^{-\frac{\delta}{1+\delta}}-1\right)^{-\frac{1+\delta}{\delta}},
\end{align*}
which is independent of $w$ and is the distribution function of a bivariate Clayton copula with parameter $\frac{\delta}{1+\delta}$.\\
Now it remains to calculate the differential in \autoref{eq:Joe1997} and plug in the conditional distributions:
\[
\frac{\partial}{\partial \xi_2}C_{U,V;W}(\xi_1,\xi_2;w)=\left(\xi_1^{-\frac{\delta}{1+\delta}}+\xi_2^{-\frac{\delta}{1+\delta}}-1\right)^{-\frac{1+\delta}{\delta}-1}\xi_2^{-\frac{\delta}{1+\delta}-1}.
\]
Finally,
\begin{align*}
C_{U|V,W}(u|v,w)&=\left\{\left(v^{-\delta}+w^{-\delta}-1\right)w^{\delta}+\left(u^{-\delta}+w^{-\delta}-1\right)w^{\delta}-1\right\}^{-\frac{1+2\delta}{\delta}}\\
&\quad\cdot\left(v^{-\delta}+w^{-\delta}-1\right)^{\frac{1+\delta}{\delta}\cdot\frac{1+2\delta}{1+\delta}}w^{(-\delta-1)\cdot\frac{-1-2\delta}{1+\delta}}\\
&=w^{-(1+2\delta)}\left(v^{-\delta}+w^{-\delta}-1+u^{-\delta}+w^{-\delta}-1-w^{-\delta}\right)^{-\frac{1+2\delta}{\delta}}\\
&\quad\cdot\left(v^{-\delta}+w^{-\delta}-1\right)^{\frac{1+2\delta}{\delta}}w^{1+2\delta}\\
&=(u^{-\delta}+v^{-\delta}+w^{-\delta}-2)^{-\frac{1+2\delta}{\delta}}\cdot (v^{-\delta}+w^{-\delta}-1)^{\frac{1+2\delta}{\delta}}.
\end{align*}
Inversion with respect to $u$ gives us the sought after conditional quantile function of a three-dimensional Clayton copula: 
\[
C^{-1}_{U|V,W}(\alpha|v,w)=\left\{(\alpha^{-\frac{\delta}{1+2\delta}}-1)(v^{-\delta}+w^{-\delta}-1)+1\right\}^{-\frac{1}{\delta}}.
\]
A visualization of the conditional median and 95\% quantile function of a three-dimensional distribution with underlying Clayton copula with parameter and marginals as specified in \autoref{tab:scenarios} is given in \autoref{fig:Clayton_quantiles}.
\end{appendix}
\section*{Acknowledgments}\label{sec:acknowledgment}
We would like to thank three anonymous referees for their comments which helped to improve the manuscript. This work was supported by the German Research Foundation (DFG grant CZ 86/4-1).

\bibliography{Vine_Regression}{}
\bibliographystyle{asa}

\end{document}